\documentclass[a4paper,11pt]{article}
\pdfoutput=1 

\usepackage{jheppub} 

\usepackage[normalem]{ulem} 

\usepackage{amsmath,amssymb,bm,mathrsfs}
\usepackage{hyperref}
\usepackage{url}
\usepackage{graphicx}
\usepackage{mathtools}
\usepackage{lipsum}
\usepackage[table]{xcolor}
\usepackage{multirow}
\usepackage{colortbl}
\usepackage{adjustbox}

\definecolor{JungleGreen}{cmyk}{0.99,0,0.52,0}

\preprint{CERN-TH-2024-195}

\title{ Wormhole-Induced ALP Dark Matter }

\author[a]{Dhong Yeon Cheong,}
\author[b, c]{Koichi Hamaguchi,}
\author[b]{Yoshiki Kanazawa,}
\author[d *]{Sung Mook Lee,}
\author[b]{Natsumi Nagata,}
\author[a,e *]{Seong Chan Park \note[*]{correspondence}}

\affiliation[a]{Department of Physics and IPAP, Yonsei University, Seoul 03722, Republic of Korea}
\affiliation[b]{Department of Physics, University of Tokyo, Tokyo 113--0033, Japan}
\affiliation[c]{ Kavli Institute for the Physics and Mathematics of the Universe (Kavli
IPMU), University of Tokyo, Kashiwa 277--8583, Japan}
\affiliation[d]{Theoretical Physics Department, CERN, CH-1211 Gen\`eve 23, Switzerland}
\affiliation[e]{Korea Institute for Advanced Study, Seoul 02455, Republic of Korea}

\emailAdd{dhongyeon@yonsei.ac.kr}
\emailAdd{hama@hep-th.phys.s.u-tokyo.ac.jp}
\emailAdd{kanazawa@hep-th.phys.s.u-tokyo.ac.jp}
\emailAdd{sungmook.lee@cern.ch}
\emailAdd{natsumi@hep-th.phys.s.u-tokyo.ac.jp}
\emailAdd{sc.park@yonsei.ac.kr}

\abstract{
Non-perturbative gravitational effects induce explicit global symmetry breaking terms within axion models.
These exponentially suppressed terms in the potential give a mass contribution to the axion-like particles (ALPs). 
In this work we investigate 
this scenario with a scalar field charged under a global $U(1)$ symmetry and having a non-minimal coupling to gravity.
Given the exponential dependence, the ALP can retain a mass spanning a wide range, which can act as a dark matter component.
We specify pre-inflationary and post-inflationary production mechanisms of these ALPs, with the former from the misalignment mechanism and the latter from both the misalignment and cosmic-string decay.
We identify the allowed parameter ranges that explain the dark matter abundance for both a general inflation case and a case where the radial mode scalar drives inflation, each in metric and Palatini formalisms.
We show that the ALP can be the dominant component of the dark matter in a wide range of its mass, $m_{a} \in [10^{-21}~\mathrm{eV},\, \mathrm{TeV}]$, depending on the inflationary scenario and the $U(1)$ breaking scale.
These results indicate that ALPs can be responsible for our dark matter abundance within a setup purely from non-perturbative gravitational effects.
}

\begin{document} 
\maketitle
\flushbottom

\section{Introduction}

%
Among beyond the Standard Model (BSM) theories, axion or axion-like particles (ALPs) have been obtaining much attention due to their unique characteristics.%
\footnote{In this work, the term `axion' refers to the QCD axion which is particularly motivated to solve the string CP problem and have a designated coupling to Standard Model fields. 
Other than that, we will use the terminology `ALP'. 
}
First of all, the existence of the axion has the potential to address two major unsolved problems in particle physics and cosmology, namely the strong CP problem~\cite{Peccei:1977hh,Peccei:1977ur, Wilczek:1977pj, Weinberg:1977ma} and the nature of dark matter~\cite{Preskill:1982cy, Abbott:1982af, Dine:1982ah}.
For the latter case, frameworks that generalize beyond the QCD axion can also provide good dark matter candidates. These ALP models are generally associated with global $U(1)$ symmetries, where the ALP mass is determined by the symmetry-breaking terms and can span a much wider range.
In general, at the very least, all global symmetries are expected to be explicitly broken in the presence of gravity~\cite{Kallosh:1995hi,Banks:2010zn,Witten:2017hdv,Harlow:2018jwu,Harlow:2018tng}. 
One definite origin of the symmetry breaking is coming from non-perturbative gravitational effects, which include gravitational instantons represented by Euclidean wormhole solutions~\cite{Lee:1988ge, Giddings:1989bq, Abbott:1989jw, Coleman:1989zu, Kallosh:1995hi, Hebecker:2016dsw,  Alonso:2017avz, Hertog:2018kbz, Hebecker:2018ofv, Loges:2022nuw, Andriolo:2022rxc, Loges:2023ypl,Jonas:2023ipa, Kanazawa:2023xzy, Martucci:2024trp, Hertog:2024nys}.
For the QCD axion scenario, this explicit PQ symmetry breaking becomes another source in addition to the QCD instanton effects, which shifts the vacuum expectation value (vev) away from the desired value that solves the strong CP problem, leading to the `axion quality problem'~\cite{Dine:1986bg,Kamionkowski:1992mf, Barr:1992qq, Holman:1992us, Ghigna:1992iv}.
However, the possible implications of gravitational global symmetry breaking reach out to any Goldstone bosons, including ALPs beyond the QCD axion.
These symmetry breaking terms can be a universal source giving the mass of general pseudo Nambu-Goldstone bosons (pNGBs)~\cite{Alonso:2017avz, Alvey:2020nyh}. 
These now massive ALPs can then be a dark matter component, in analogy to axion dark matter.
A nonzero initial angle induces a misalignment mechanism~\cite{Preskill:1982cy, Abbott:1982af, Dine:1982ah}, and cosmic strings associated with the global symmetry can also emit ALP dark matter particles~  \cite{Sikivie:1982qv,Vilenkin:1984ib,Davis:1986xc,Vincent:1996rb,Kawasaki:2014sqa,Vilenkin:1986ku}, with the mass, in this case, being solely determined through gravitational effects. 
The mass range highly depends on the wormhole action value, which itself depends on the axion model context.
Especially, in Refs.~\cite{Hamaguchi:2021mmt,Cheong:2022ikv, Cheong:2023hrj}, it was shown that a large non-minimal coupling to gravity with coupling $\xi$ significantly alters the wormhole properties with respect to minimal gravity cases. 
This motivates the necessity to rigorously revisit the possible effects of these gravitational instantons on ALP DM.
In this paper, we explore the possibility that non-perturbative gravitational effects can induce explicit global symmetry breaking in ALP models.
Specifically, we consider a global $U(1)$ scalar field non-minimally coupled to gravity, and investigate how the resulting ALP mass, determined by exponentially suppressed symmetry breaking terms, can span a wide range.
Notably, for $\xi = 0$, the wormhole-induced ALP mass resulting from explicit $U(1)$ symmetry breaking is too large to be a viable dark matter candidate. This also emphasizes the need to introduce a non-minimal coupling as a minimal mechanism to suppress the $U(1)$ global symmetry breaking caused by the wormhole.

Our study covers both pre-inflation and post-inflation production mechanisms.
We examine the allowed parameter spaces that explain the dark matter abundance, considering both general inflationary scenarios and cases where the radial mode scalar drives inflation.
Our analysis includes both the metric and Palatini formalisms, and we demonstrate that ALPs can account for the observed dark matter density over a broad range of mass and symmetry breaking scales, driven purely by non-perturbative gravitational effects.
This paper is organized as follows. 
We first review the wormhole properties within an analytic framework, and obtain the ALP mass expression in Section~\ref{section:ALP Mass from the Wormhole Solution}.
In Section~\ref{sec:ALP_DM_abundance}, we further analyze the ALP dark matter production mechanisms, considering the global symmetry breaking scale.
We then present possible parameter ranges that explain our dark matter abundance after identifying relevant constraints {for both general inflation in Section~\ref{sec:generic_inflation} and the case when the radial mode of the complex scalar being the inflaton field in Section~\ref{sec:radial_mode_inflation}}.
In Section~\ref{section:conclusion}, we conclude with possible further implications of these wormhole dark matter candidates.

\section{ 
ALP Mass from the Wormhole Solution}

\label{section:ALP Mass from the Wormhole Solution}

\subsection{Non-minimally Coupled $U(1)$ Scalar}

%
In this work, we consider a model of a complex scalar field $\Phi = \frac{1}{\sqrt{2}} \rho  e^{i \theta}$ with a $U(1)$ symmetry which is spontaneously broken by the Mexican hat potential:
\begin{equation}
    \begin{aligned}
    S
    &= \int d^4x \sqrt{|g|} \left[
    - \frac{M^2+2\xi|\Phi|^2}{2}R(\Gamma)  +|\partial_{\mu}\Phi|^2+ \lambda_{\Phi} \left(|\Phi|^2 -f_a^2/2 \right)^2 \right] \\
     &= \int d^4x \sqrt{|g|} \left[
    - \frac{M^2+\xi \rho^2}{2}R(\Gamma) + \frac{1}{2} (\partial_{\mu} \rho)^{2}  + \frac{1}{2} \rho^{2}(\partial_{\mu} \theta)^{2} + \frac{\lambda_{\Phi}}{4} \left( \rho^2 -f_{a}^{2} \right)^2 \right]
    , \label{eq:action_def}
    \end{aligned}
\end{equation}
where $g \equiv \det g_{\mu\nu}$, and the mass parameter $M$ is defined as $M^{2} \equiv M_{P}^2 - \xi f_{a}^{2}$ with the reduced Planck mass $M_{P} \equiv 1/\sqrt{8 \pi G} \approx 2.4\times 10^{18} \, {\rm GeV}$ to guarantee that the gravity sector reduces to canonical Einstein-Hilbert action with the vev of $\rho$ as $ \langle \rho \rangle = f_{a} $.
Normally, pNGB degree of freedom $\theta$ is canonicalized to $a \equiv f_{a} \theta$ and we refer this axial mode as ALP. On the other hand, we will call $ \rho $ as a radial mode or radial field.
Note that we also have included the non-minimal coupling of the $U(1)$ scalar $ \Phi $ to the Ricci scalar $R$ with coupling $\xi$.
Especially, we will refer the special case of $\xi = M_{P}^{2} / f_{a}^{2}$ as the induced gravity limit, or Gidding-Strominger (GS) limit~\cite{Giddings:1987cg,Hamaguchi:2021mmt,Cheong:2022ikv}.
Also, we intentionally denote the affine connection $\Gamma$ dependence of the Ricci scalar $R$ explicitly, while we do not specify whether $\Gamma$ is determined by the metric or not in prior.
In fact, the choice of the connection depends on the formulation of the gravity one chooses.
In this work, we consider two formulations of the gravity: metric and Palatini \cite{Einstein:1925,Ferraris:1982}.
In the metric formulation, affine connection $\Gamma$ is given by the Christoffel symbol from the first, while it is obtained from the equations of motion in Palatini formulation.
Although these two formulations are equivalent in pure Einstein-Hilbert gravity, they differ each other once one adds non-minimal coupling and so does the physical predictions \cite{Ferraris:1992dx,Magnano:1993bd}.
%

\subsection{Wormhole Solution }
\label{sec:wormhole_solution}

The numerical evaluation of the axionic wormhole action $S_{w}$ and the wormhole throat size $L_{w}$ of the model with a non-minimal coupling as given in Eq.~\eqref{eq:action_def}
was studied in Refs.~\cite{Hamaguchi:2021mmt,Cheong:2022ikv} and analytically analyzed in Ref.~\cite{Cheong:2023hrj}.
For our purpose, we generalize the results of Ref.~\cite{Cheong:2023hrj} to incorporate the case near the GS limit.\footnote{For the following equations to hold, we restrict ourselves to be in the regime where $\lambda_{\Phi} < \mathcal{O}({1}) $ , or more specifically, $Q \equiv n^2 \lambda_{\Phi}^2 / (8 \pi^4) < 1 $.}
For an action 
\begin{align} 
S = \int d^4 x \sqrt{|g|} \left( -\frac{M_P^2}{2} R 
+ \frac{1}{2}   G(\rho) (\partial_\mu\rho)^2  +\frac{1}{2}   F^2_a(\rho) (\partial_\mu\theta)^2\right),
\label{eq:action_with_GandF}
\end{align}
containing functions $G(\rho)$ and $ F_{a}(\rho) $ with the Euclidean wormhole metric {in spherical coordinates}
\begin{align}
ds^2=\frac{d r^2}{\left(1-L_{w}^4 / r^4\right)}+r^2 d \Omega_3^2
\end{align}
where $ d\Omega_{3}^{2} $ is the metric of the unit 3-sphere, the wormhole throat size $L_{w}$ is given by
\begin{align} 
    L_{w}^2 = \frac{n }{2\pi^2 \sqrt{6} } \left( \frac{1}{M_{P} F_{a} (\rho_{0}) } \right) 
\end{align}
with the radial mode field value $\rho_{0}$ at the wormhole throat obtained {from the condition}
\begin{align}
\int_{\rho_{\infty}}^{\rho_0} \frac{d\varrho}{M_P}\,  \frac{\sqrt{G(\varrho)}}{\sqrt{F^2_a(\rho_0)/F^2_a(\varrho) -1}} \equiv K(\rho_{0}, \rho_{\infty}) =   
\frac{\pi \sqrt{6}}{4} 
\label{eq:actionKfunction}
\end{align}
where $\rho_{\infty}$ is the vev of the $\rho$ field at $r\rightarrow \infty$.
Then, the wormhole action is obtained as
\begin{align}
    S_{w}(\rho_{0}, \rho_{\infty} ) & = n \int_{\rho_{\infty}}^{\rho_{0}} \frac{d\varrho}{F_{a}(\varrho) } \frac{\sqrt{G(\varrho)}}{\sqrt{1 - F_{a}^{2}(\varrho) / F_{a}^{2}(\rho_{0})}} \, .
\label{eq:action_integral}
\end{align}
For the specific case of Eq.~(\ref{eq:action_def}), after performing the redefinition of the metric $g_{\mu\nu} \rightarrow \Omega^{2} g_{\mu\nu} $ with
$ \Omega^{2} \equiv (M^{2} + \xi \rho^{2})/M_{P}^{2} $ and identifying $\rho_{\infty} = f_{a}$, the functions $G(\rho)$ and $F_{a} (\rho) $ are given as 
\begin{align}
    G (\rho) = \frac{1+\xi \rho^2 (1+\alpha \xi) - \xi f_{a}^2 }{(1+\xi(\rho^2 - f_{a}^2 ))^2} \, , && F_{a}( \rho ) = \frac{\rho }{\sqrt{1+\xi(\rho^2 - f_{a}^2)}} 
\end{align}
where $\alpha=6$ and 0 for metric and Palatini formalisms, respectively~\cite{Cheong:2023hrj}.
Inserting these expressions into Eq.~(\ref{eq:actionKfunction}) and Eq.~(\ref{eq:action_integral}) yields
\begin{align}
    K(\rho_{0}, f_{a} ) &= \sqrt{\frac{(1+\alpha \xi)(1 - \xi f_{a}^2 + \xi \rho_{0}^2) }{\xi (1 - \xi f_{a}^2)}} \arccos\left[ \sqrt{\frac{1+ \alpha \xi^2 f_{a}^2}{1- \xi f_{a}^2 + (1+\alpha\xi)\xi \rho_{0}^2 }}\right] \nonumber \\
    & \hspace{4cm} - \sqrt{\alpha } \arctan\left[\sqrt{\frac{(1- \xi f_{a}^2) \xi^2 \alpha (\rho_{0}^2 - f_{a}^2 )}{(1+\alpha \xi^2 f_{a}^2 )(1 - \xi f_{a}^2 + \xi \rho_{0}^2 )}} \right]
\end{align}
and
\begin{equation}
\begin{aligned}
    S_{w} (\rho_{0}, f_{a} ) &= \sqrt{\frac{\xi( 1+ \alpha \xi) }{1 - f_{a}^2 \xi }} \rho_{0} \arccos\left[ \sqrt{\frac{1 + \xi^2 \alpha f_{a}^2 }{1 - \xi f_{a}^2 + \xi(1 + \alpha \xi) \rho_{0}^2 }}\right] \\
    & + \frac{1}{2} \left( \ln \left[ {1 + \sqrt{\frac{(1- \xi f_{a}^2 )(\rho_{0}^2 - f_{a}^2 )}{(1 + \xi^{2} \alpha f_{a}^2) \rho_{0}^2}}} \right] - \ln \left[ {1 -\sqrt{\frac{(1- \xi f_{a}^2 )(\rho_{0}^2 - f_{a}^2 )}{(1 + \xi^{2} \alpha f_{a}^2) \rho_{0}^2}}} \right] \right) \,.
\end{aligned}
\end{equation}
Different $\xi$ dependence in the $S_{w}$ and $L_{w}$ appears with the $\xi$ value range of interest.
For small non-minimal couplings, $\xi \ll 1$, the wormhole throat size becomes effectively independent of the $\xi$ value. In both gravity formalisms, the wormhole has a throat of Planck length, and the wormhole action is suppressed logarithmically:
\begin{align}
S_{w}^{\xi \ll 1} \simeq n \ln \left( \frac{M_{P}}{f_{a} } \right) \, , && (L_{w}^{\xi \ll 1 })^{2} \simeq \frac{ n }{3\pi^{3} M_P^2 }
\end{align}
where $n$ is the number of charge \cite{Abbott:1989jw, Kallosh:1995hi}.
In fact, for the QCD axion, this smallness of the wormhole action leads to unsuppressed $U(1)$ symmetry breaking terms which could be relevant when discussing the QCD axion quality problem.
On the other hand, for values $1 \ll \xi \lesssim M_{P}^{2} / f_{a}^{2} $, the wormhole action is dominated by the throat contribution yielding \cite{Cheong:2023hrj}
\begin{align}
    S_{w}^{\xi \gg 1} \simeq 
    \begin{dcases}
      \frac{\pi\sqrt{30}}{4} n  \xi^{1/2} & (\text{metric})  \\
      \frac{\pi\sqrt{6}}{4} n \xi^{1/2} & (\text{Palatini})
    \end{dcases}, &&
     (L_{w}^{\xi \gg 1})^{2} \simeq 
    \begin{dcases} 
           \frac{\sqrt{3}n \xi^{1/2} }{2\pi^2 \sqrt{10} M_P^{2}}   & (\text{metric}) \\
           \frac{ n \xi^{-1/2} }{2\pi^2 \sqrt{6} M_P^{2}}   & (\text{Palatini})
    \end{dcases}.
\end{align}
The limiting case is when $\xi$ value takes its maximum value, $\xi = M_{P}^2/f_{a}^{2} $, which becomes equivalent to the GS wormhole, with the wormhole action fixed as~\cite{Giddings:1987cg,Alonso:2017avz,Hamaguchi:2021mmt,Cheong:2022ikv}
\begin{align}  
S_{w }^{\text{GS}} = \frac{ \sqrt{6} \pi n M_{P}}{ 4 f_{a}} \, , &&(L_{w}^{\text{GS}})^{2} = \frac{ n }{2\pi^2 \sqrt{6}  M_{P} f_{a} } \, .
\end{align}
Note that this result does not depend on the formulations of gravity.
Henceforth, we always set $n=1$ because this gives the leading contribution.%
\footnote{Here, we do not take into account the Gibbons-Hawking-York (GHY) term. When there is a boundary of the manifold in the consideration, the GHY term should be introduced in the case of the metric formulation to make the variation principle well-defined \cite{York:1972sj,Gibbons:1976ue}.
In this case, the action value is suppressed by a factor $(1-2/\pi)$ multiplied where the second term is from the GHY term \cite{Alonso:2017avz}.}
%

\subsection{ALP Mass}

On dimensional grounds, the $U(1)$ symmetry breaking terms in the potential have the following schematic form
\begin{align}
\Delta V  \sim   e^{-S_{w}} L_{w}^{-3} (\Phi + \Phi^{*}) 
+ \cdots 
 \sim e^{-S_{w}} \frac{\rho}{L_{w}^{3}} \cos \left( \frac{a}{f_{a}} \right)  + \cdots 
\end{align}
with $ \cdots $ representing subleading order contributions associated with larger charges. 
The exponential contribution of the wormhole action clearly shows its non-perturbative nature.
Asymptotically, the heavy $\rho$ field value stabilizes to have a vacuum expectation value $\langle \rho \rangle = f_{a}$ giving a mass to the ALP in the form 
\begin{align}
    m_{a}^{2} \sim \frac{1}{f_{a} L_{w}^{3}} e^{-S_{w}} \label{eq:ALPmasswh}
\end{align}
assuming there is no other source of the symmetry breaking that would induce additional mass contributions.\footnote{Hereafter, we assume $m_a^2\lesssim m_\rho^2 \simeq 2\lambda_\Phi f_a^2$, for simplicity.
In the case of $m_a^2\gg m_\rho^2$, there are corrections to the vev of the radial mode and the ALP mass as 
$\langle \rho \rangle \to (m_a/m_\rho)^{2/3} \langle \rho \rangle$ 
and $m_a^2 \to (m_a/m_\rho)^{2/3} m_a^2$, but the following discussion does not change qualitatively.}
In this sense, unless there is a large cancellation between symmetry breaking terms, this mass in Eq.~(\ref{eq:ALPmasswh}) may be regarded as the ALP's minimal mass originating from the gravitational coupling.
For the QCD axion, restrictions on the wormhole action coming from the axion quality problem also need to be considered. 
For the QCD axion quality to be guaranteed, the wormhole action needs to be sufficiently large, $S_{w} \gtrsim 190$, which suppresses the wormhole mass contribution to negligible values compared to the 
original predictions of the QCD axion mass $ m_{a} \sim \Lambda_{\rm QCD}^{2} / f_{a}$ with $ \Lambda_{\rm QCD}$ is the QCD confinement scale.
Therefore, we do not consider the QCD axion case in this work and take the ALP to be a generic $U(1) $ pNGB. 
\begin{figure*}
    \centering
    \includegraphics[width=0.45\textwidth]{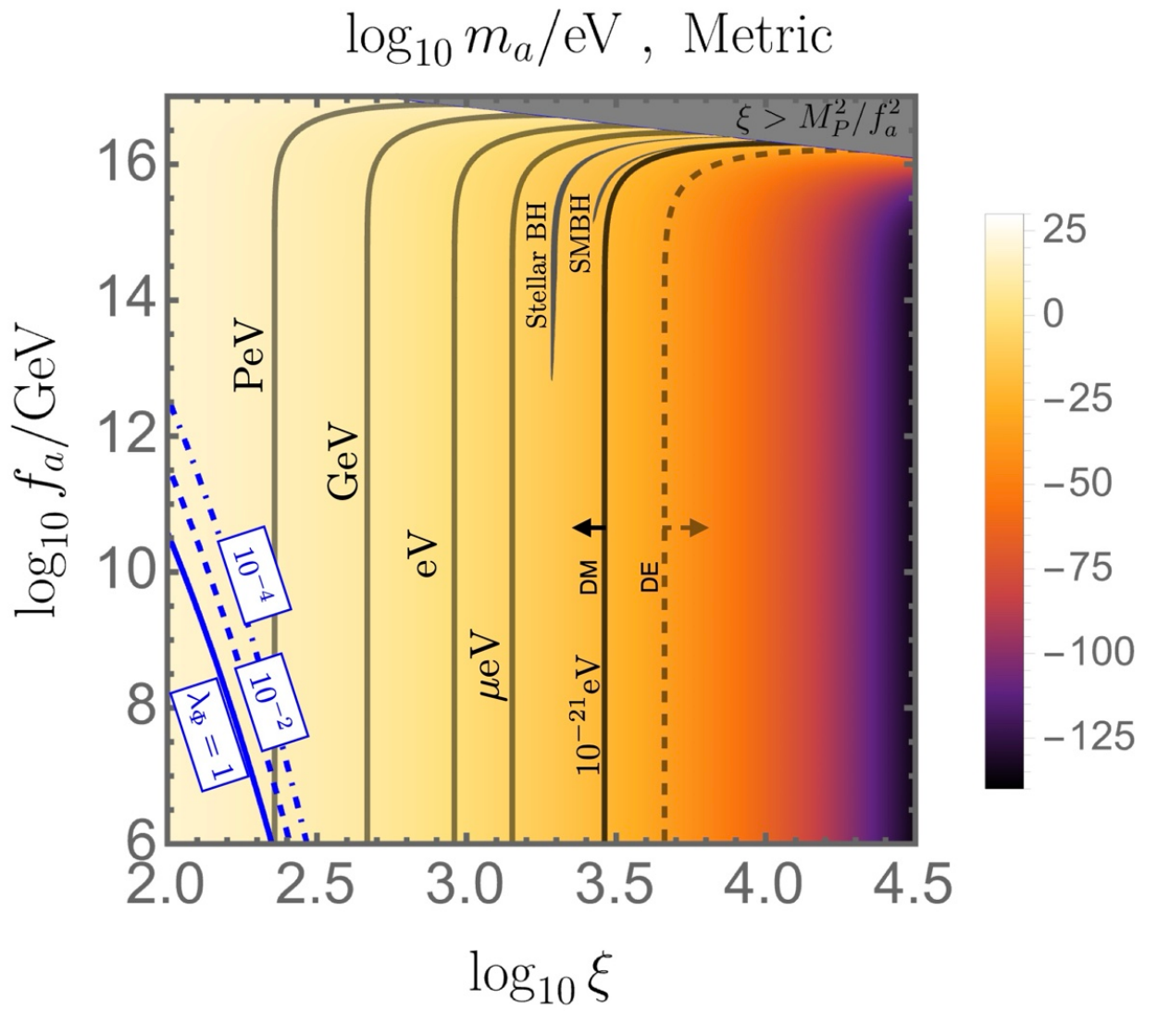}
    \hspace{1cm}
    \includegraphics[width=0.45\textwidth]{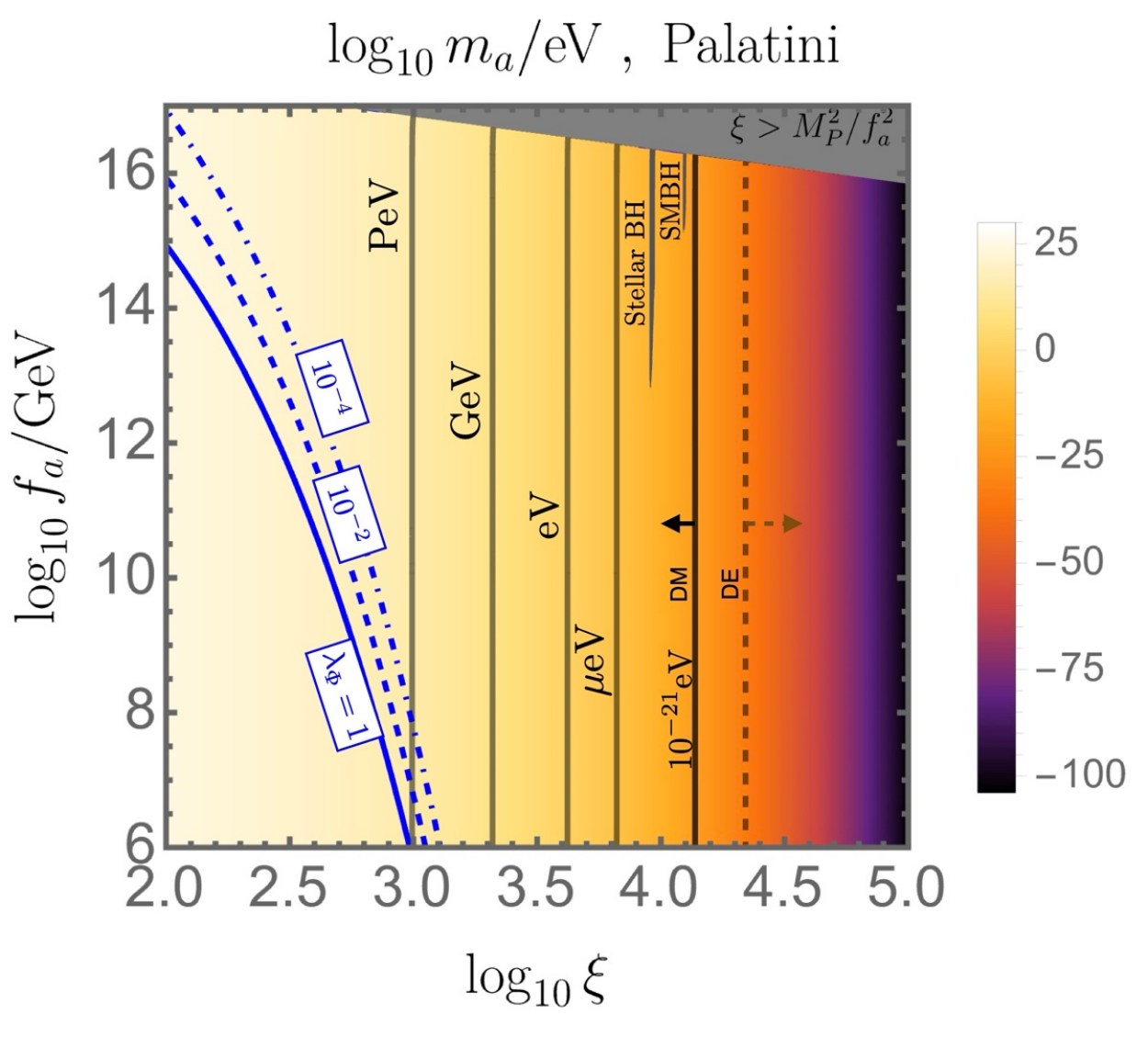}
    \caption{
    The wormhole-induced ALP mass in terms of $(\log_{10} \xi,~ \log_{10} {f_{a} }/{\text{GeV}})$ for metric (left) and Palatini (right) formulations.
    The black solid line corresponds to the lower bound of fuzzy dark matter $m_{a} \simeq 10^{-21}\,\text{eV}$~\cite{Ferreira:2020fam,Workman:2022ynf,OHare:2024nmr}, while the gray dashed line corresponds to the upper bound for dark energy candidates $ m_{a} \simeq H_0 \simeq 10^{-33} \,\text{eV}$.
    Blue lines denote the boundaries where $ m_{\rho} = m_{a} $ depending on values of $\lambda = (1, 10^{-2}, 10^{-4})$ with solid, dashed, dot-dashed lines respectively. Below these lines, the radial mode mass is smaller than the ALP mass.
    Gray shaded regions labeled Stellar BH and SMBH are constraints from superradiance for stellar mass BH and SMBH respectively \cite{AxionLimits, Stott:2018opm, Baryakhtar:2020gao, Hoof:2024quk}. See the main text for details.
    }
    \label{fig:ALPmass}
\end{figure*}
Figure~\ref{fig:ALPmass} depicts the ALP mass given in Eq.~\eqref{eq:ALPmasswh} for both metric and Palatini formulation of gravity as a function of the axion decay constant $f_{a}$ and the non-minimal coupling to gravity $\xi$.
In both cases, the gray region is excluded as in this parameter space the coefficient of the Ricci scalar becomes negative with $ \xi > M_{P}^{2} / f_{a}^{2} $.
For a fixed $f_{a}$, a larger $\xi$ implies a larger wormhole action $S_{w}$ giving a smaller ALP mass.
While the mass hardly depends on $f_{a}$ in the Palatini case, for the metric formalism there is a rapid change to the GS limit at larger $f_{a}$ values. In GS limit, both formulations give the same mass as expected.
When these ALPs acts as DM candidates, these may be subject to additional constraints, including isocurvature perturbations (depending on its generation mechanism) and structure formation.
We will revisit isocurvature constraints in the next section, while successful structure formation demands a model independent bound $ m_{a} \gtrsim 10^{-21} \, \text{eV} $ \cite{Ferreira:2020fam,Workman:2022ynf,OHare:2024nmr}. 
Ultralight bosons are also subject to constraints from black hole superradiance.
Recent studies on stellar mass black holes and supermassive black holes put constraints on the ALP mass depending on the axion decay constant $f_{a}$~\cite{AxionLimits, Stott:2018opm, Baryakhtar:2020gao, Hoof:2024quk}.
We depict the masses $ 8.6\times 10^{-20}~\mathrm{eV} \lesssim m_{a} \lesssim 5.6\times 10^{-19}~\mathrm{eV}  $ corresponding to supermassive black hole superradiance, and $ 1.9\times 10^{-13}~\mathrm{eV} \lesssim m_{a} \lesssim 2.7\times 10^{-12}~\mathrm{eV}  $ for stellar mass black hole superradiance in Figure~\ref{fig:ALPmass}.
On the other hand, these ALPs can behave as dark energy (DE) for ALP masses smaller than the current universe's Hubble parameter $H_{0} \simeq  10^{-33} \, {\rm eV}$.
Recently, these ALP DE scenarios particularly gained increased interest (i) as an early dark energy component being a possible solution to the Hubble tension \cite{Kamionkowski:2022pkx} or/and (ii) being responsible for the observed isotropic CMB birefringence through the Chern-Simon's coupling with photons \cite{Komatsu:2022nvu}. We also note that such time-varying dark energy has received much attention due to the
recent DESI results \cite{DESI:2024mwx}.
In order for wormhole-induced ALPs to be dark energy, given the small $H_{0}$ value, we need a large non-minimal coupling $\xi$; $ \xi \gtrsim 3.6 \times 10^{3} $ for the metric case (except near the GS limit), and $ \xi \gtrsim 4.3 \times 10^{4} $ for the Palatini case. From the next section, we focus on the possibility of ALP DM in more detail.


\section{ALP Dark Matter Abundance}
\label{sec:ALP_DM_abundance}

Having the relations for the ALP mass induced by the wormhole solution, we revisit the parameter regimes of ALP DM  associated with a large non-minimal coupling to gravity.
In general, the ALP DM abundance strongly depends on the relative magnitudes of the $U(1)$ breaking scale $f_a$, reheating temperature $T_{\rm reh}$, and the de-Sitter (dS) temperature during inflation $T_{\rm dS} \equiv H_{\rm inf}/(2\pi)$ quantifying the amount of the quantum fluctuation of the massless particle during the inflation, where $ H_{\rm inf}$ is the Hubble scale during the inflation.
For instance, if $ T_{\rm dS} > f_{a}$, the amplitude of the quantum fluctuation dominates over the vev, and the PQ symmetry remains unbroken during the inflation. In the subsequent subsections, we discuss each of the following cases separately in more detail.
\begin{enumerate}
\item $f_a > \max\{ T_{\rm reh}, T_{\rm dS} \}$: pre-inflationary scenario. (Section~\ref{subsec:pre-inf}.)
\item $f_a < \max\{ T_{\rm reh}, T_{\rm dS} \}$: post-inflationary scenario.
(Section~\ref{subsec:post-inf}.)
\end{enumerate}
We summarize the various cases of ALP DM abundance depending on the hierarchy between axion decay constant ($ f_{a} $), the Hubble scale of the inflation ($H_{\rm inf}$), at the end of the reheating ($H_{\rm reh}$), and when the coherent oscillation of the ALP starts ($H_{\rm osc}$) in Table~\ref{table:ALP-misalignemtn-scenarios}.
%

\begin{table}[t]
\centering
\begin{adjustbox}{width=0.8\textwidth,center=\textwidth}
\begin{tabular}{|c|c|}
\hline & $f_a>T_{\mathrm{reh}}$ \\
\hline $f_a>T_{\mathrm{dS}}$ 
& 
\cellcolor{orange!10} 
\begin{tabular}{c} 
\mbox{}\\
\textcolor{orange}{\bf ``Pre-Inflationary Scenario" $\left( \theta_i \in[-\pi, \pi] \right)$} [Section~\ref{subsec:pre-inf}] \\
$\displaystyle{\begin{cases}
H_{\mathrm{osc}} \lesssim H_{\mathrm{reh}} & \text{osc. during RD at $ H = H_{\rm osc} $ [Eq.~\eqref{eq:pre_1}]}
\\
H_{\text {reh }} \lesssim H_{\mathrm{osc}} \lesssim H_{\mathrm{inf}} & \text{osc. during reheating at $ H = H_{\rm osc}$ [Eq.~\eqref{eq:pre_2}]}
\\
H_{\mathrm{osc}} \gtrsim H_{\mathrm{inf}} & \text{no misalignment}
\end{cases}}$\\
\mbox{}
\end{tabular} 
 \\
\hline 
$f_a<T_{\mathrm{dS}}$ & 
\cellcolor{blue!10}
\begin{tabular}{c} \mbox{} \\ \textbf{Case II} [Section~\ref{subsubsection:post_2}]
\\
$\displaystyle{\begin{cases}
H_{\mathrm{osc}} \lesssim H_{\text {reh }} & \text{osc. during RD at $ H = H_{\rm osc} $ [Eq.~\eqref{eq:pre_1}]}
\\
H_{\mathrm{osc}} \gtrsim H_{\text {reh }} & \text{osc. during reheating at $ H= H_{\rm osc} $ [Eq.~\eqref{eq:pre_2}]}
\end{cases}}$\\
\mbox{}  \\
\end{tabular} \\
\hline
\hline \multicolumn{2}{|c|}{$f_a<T_{\mathrm{reh}}$ }   \\
\hline  
\multicolumn{2}{|c|}{
\cellcolor{blue!10} 
\begin{tabular}{c} 
\\
\textbf{Case I} [Section~\ref{subsubsection:post_1}]
\\
$\displaystyle{\begin{cases}
H_{\mathrm{osc}} \lesssim H_c & \text{osc. during RD at $ H = H_{\rm osc} $ [Eq.~\eqref{eq:pre_1}]}
\\
H_{\mathrm{osc}} \gtrsim H_c & \text{osc. during RD at $H=H_c$ [Eq.~\eqref{eq:post_2}]}
\end{cases}}$ 
\mbox{}
\end{tabular}}\\
 \multicolumn{2}{|c|}{\cellcolor{blue!10} 
\begin{tabular}{c} \mbox{} \\
\textcolor{blue}{\bf ``Post-Inflationary Scenario" $ \left( \left\langle\theta_i^2\right\rangle=\pi^2 / 3 \right) $ }\\
\mbox{} \\
\end{tabular}} \\
\hline 
\end{tabular}
\end{adjustbox}
\caption{Cases of ALP DM misalignment scenarios.
The orange colored box corresponds to pre-inflationary scenarios, while blue colored ones are for post-inflationary scenarios. 
}
\label{table:ALP-misalignemtn-scenarios}
\end{table}

%
In the following, we parameterize the reheating temperature as
\begin{align}
    T_{\rm reh} = \epsilon_{\rm eff} \sqrt{H_{\rm inf} M_{P}}
\end{align}
where $\epsilon_{\rm eff}$ denotes the efficiency of the reheating, which is in principle determined by the couplings of the inflaton field with Standard Model and DM particles.
It takes the maximal value in the limit of instantaneous reheating:
\begin{align}
\epsilon_{\mathrm{eff}} \le \epsilon_{\mathrm{eff}}^{\rm max} = \left( \frac{90}{g_*(T_{\rm reh}) \pi^2}\right)^{1/4},
\end{align}
where $ g_*(T_{\rm reh})$ is the effective number of {the relativistic} degrees of freedom at the end the reheating. With this definition, $H_{\rm reh}$ can be expressed as $H_{\rm reh} = (\pi^2 g_{*}(T_{\rm reh})/90)^{1/2} \epsilon_{\rm eff}^2 H_{\rm inf}$.
Another important parameter to determine the ALP abundance is the Hubble parameter when the ALP starts its coherent oscillation. In the quadratic ALP potential, $V(a)\simeq m_a^2 a^2/2$, the ALP starts to oscillate when the Hubble parameter becomes comparable to the ALP mass, $H=H_{\rm osc}\simeq m_a/3$.
%

\subsection{$f_a > \max\{ T_{\rm reh}, T_{\rm dS} \}$: Pre-inflationary Scenario}
\label{subsec:pre-inf}

If $ f_{a} > \max\{ T_{\rm reh}, T_{\rm dS} \} $, $U(1)$ symmetry is spontaneously broken during the inflation and the symmetry is never restored after that.
In this case, unless the ALP is heavier than the Hubble scale of the inflation so that $H_{\rm osc} \,{ \gtrsim }\, H_{\rm inf}$, the initial ALP field value {of the observable universe} is randomly determined by {a single} initial misalignment angle $\theta_i \in [-\pi,\pi]$. With the quantum fluctuation during the inflation on top of this classical misalignment angle, the averaged initial angle over several Hubble patches is given by
\begin{align}
   \langle \theta_i^{2} \rangle  =  \theta_i^{2} + \left( \frac{H_{\rm inf}}{2\pi f_{a}} \right)^{2} \, .
\end{align}
We can further divide into three cases, based on the epoch in which the ALP starts to oscillate, giving the misalignment mechanism. 

\begin{itemize}

\item $H_{\rm osc} \lesssim H_{\rm reh}$

If the ALP is sufficiently light, it stays at its initial value during reheating and begins to oscillate afterwards {during the radiation dominant universe}, at $H = H_{\rm osc} $.
The ratio of the ALP energy density ($\rho_a$) to the entropy density ($s$) at the time of the oscillation is {a conserved quantity} given by
\begin{equation}
    \begin{aligned}
         \left. \frac{\rho_{a}}{s}  \right\vert_{\rm osc} & = \left. \frac{\rho_{\rm rad}}{s}  \right\vert_{\rm osc} \cdot   \left. \frac{\rho_{a}}{\rho_{\rm rad}}  \right\vert_{\rm osc}   {\simeq} \frac{3}{4} T_{\rm osc}  \frac{g_{*}(T_{\rm osc})}{g_{*,s}(T_{\rm osc})} \cdot  \frac{ \langle \theta_{i}^{2} \rangle  f_{a}^{2} m_{a}^{2}/2}{3M_{P}^{2} H_{\rm osc}^{2}} 
\label{eq:pre_misalignment}
    \end{aligned}
\end{equation}
where $\rho_{\rm rad}$ is the radiation energy density and we have used
\begin{equation}
    \begin{aligned}
        &\frac{\rho_{\rm rad}}{s} = \frac{3}{4}T\frac{g_{*}(T)}{g_{*,s}(T)} \, ,
        \qquad
        \rho_{a} \vert_{\rm osc}\simeq \frac{1}{2}\langle \theta_{i}^{2} \rangle  f_{a}^{2} m_{a}^{2} \, , \\
        &\rho_{\rm rad} \vert_{\rm osc} = \frac{\pi^2}{30}g_*(T_{\rm osc})T_{\rm osc}^4 \simeq 3M_P^2 H_{\rm osc}^2 \, .
        \label{eq:rhooscexpression}
    \end{aligned}
\end{equation}
As the ratio $\rho_a/s$ stays constant until the present Universe, {we obtain} the final ALP DM abundance as
\begin{align}
\Omega_a h^2 
&= \frac{\rho_a/s}{(\rho_{\rm crit}/s)_0} h^2 
\simeq 
0.12 
    \langle \theta_{i}^{2} \rangle 
    \left( \frac{f_{a}}{7 \times 10^{16} \, \text{GeV}} \right)^2 \sqrt{\frac{m_{a}}{10^{-21} \, \text{eV}}} \,
\label{eq:pre_1}
\end{align}
where we used $(\rho_{\rm crit}/s)_0 \simeq 3.64 \times 10^{-9} h^2$ GeV~\cite{ParticleDataGroup:2024cfk} and assumed $ g_{*}(T_{\rm osc}) = g_{*,s}(T_{\rm osc}) = 106.75$ for simplicity.

\item $ H_{\rm reh} \lesssim  H_{\rm osc} \lesssim H_{\rm inf}$ 

In this case, the ALP oscillation starts during reheating. We assume that the equation of state {$ w \equiv p / \rho $ with the pressure $p$ and the energy density $ \rho $} during the reheating stage is parameterized by a constant $w_{\rm reh} $. Then, the ratio $\rho_a/s$ at the end of the reheating is given by
\begin{equation}
\begin{aligned}
    \left. \frac{\rho_{a}}{s}  \right\vert_{\rm reh}
&  = \left. \frac{\rho_{\rm rad}}{s}  \right\vert_{\rm reh} \cdot   \left. \frac{\rho_{ a}}{\rho_{\rm \phi}}  \right\vert_{\rm reh} = \left. \frac{\rho_{\rm rad}}{s}  \right\vert_{\rm reh} \cdot   \left. \frac{\rho_{ a}}{\rho_{\rm \phi}}  \right\vert_{\rm osc} \cdot \left( \frac{a_{\rm osc}}{a_{\rm reh}}\right)^{-3w_{\rm reh}} \\
& = \frac{9}{8} \epsilon_{\rm eff} \frac{g_{*}(T_{\rm reh})}{g_{*,s}(T_{\rm reh})}\frac{ \langle \theta_{i}^{2} \rangle f_{a}^{2} H_{\rm inf}^{1/2}}{M_{P}^{3/2}} \left( \frac{\pi \sqrt{g_{*}(T_{\rm reh})}}{9\sqrt{10}} \cdot \epsilon_{\rm eff}^{2} \frac{H_{\rm inf}}{m_{a}}\right)^{\frac{-2w_{\rm reh}}{1 + w_{\rm reh}}} \label{eq:pre_2}
\end{aligned}
\end{equation}
with $\rho_{\rm rad}/s|_{\rm osc}$, $\rho_{a} |_{\rm osc}$ taking the expression from Eq.~(\ref{eq:rhooscexpression}), and 
$\rho_\phi|_{\rm osc} = 3M_{P}^2 H_{\rm osc}^2 $ denoting the energy density of the inflaton at the period of oscillation.
In the case of $ w_{\rm reh}  = 1/3 $, there is no distinction between reheating and the subsequent radiation dominated universe, and Eq.~\eqref{eq:pre_2} reduces to Eq.~\eqref{eq:pre_misalignment}. 

\item $ H_{\rm osc } \gtrsim H_{\rm inf} $

In this case, the ALP field settles down to the potential minimum during the inflation, and there is no misalignment. Quantum fluctuations are also exponentially suppressed.

\end{itemize}

In the pre-inflationary scenario, the isocurvature perturbation can give a stringent constraint. The amount of the isocurvature perturbation strongly depends on whether the radial mode of the $U(1)$ breaking field $\rho$ is identified with the inflaton field or not. We discuss it in Section~\ref{sec:generic_inflation} and Section~\ref{sec:radial_mode_inflation}.
%

\subsection{$f_a < \max\{ T_{\rm reh}, T_{\rm dS} \}$: Post-inflationary Scenario}
\label{subsec:post-inf}

For $f_{a} $ values satisfying this criteria, the $U(1)$ symmetry is restored,
thus
giving an averaged initial misalignment value of $\langle \theta_{i}^{2} \rangle = \frac{1}{2\pi} \int_{-\pi}^{\pi} d\theta \, \theta^{2} = \pi^{2}/3$.%
\footnote{
In fact, a more precise value would depend on the exact shape of the potential at large $ \theta$ values near the maximum by having anharmonic corrections.
For instance, in the case of a cosine potential like the QCD axion, $ \langle \theta_{i}^{2} \rangle = 1.41 \cdot \pi^{2}/3$ \cite{GrillidiCortona:2015jxo,OHare:2024nmr}.
Although we take $ \pi^{2} / 3 $ as a reference value for simplicity neglecting the exact form of the potential, we do not expect large deviation from this value.}
We can further specify two separate cases within this regime,
$f_{a} \lesssim T_{\rm reh}$ and $T_{\rm reh} < f_{a} < T_{\rm dS}$. We discuss these cases in Section~\ref{subsubsection:post_1} and Section~\ref{subsubsection:post_2}, respectively.

An important consequence of the post-inflationary scenario is the formation of the cosmic string caused by the spontaneous $U(1)$ breaking~\cite{Davis:1986xc,Harari:1987ht,Lyth:1991bb}.
The decay of the cosmic string produces ALPs which also contribute to the dark matter relic.%
\footnote{The evolution of the cosmic string networks are relevant, and we note that this is currently being actively investigated~\cite{Klaer:2017qhr,Gorghetto:2018myk,Kawasaki:2018bzv,Martins:2018dqg,Buschmann:2019icd,Hindmarsh:2019csc,Hook:2018dlk,Benabou:2023ghl, Drew:2023ptp, Benabou:2023npn, Saikawa:2024bta, Kim:2024wku}.}
Due to its large theoretical uncertainties, we parameterize the amount of ALP DM from the decay of the string network with respect to the ones from misalignment as
\begin{align}
    \Omega^{\rm str}_{a} = \delta_{\rm dec} \Omega_{a}^{\rm mis}
\end{align}
where we will take $\delta_{\rm dec} \in [0, 10^{3}]$ and $ \Omega_{a}^{\rm mis} $ is {the ALP abundance from misalignment} with $ \langle \theta_{\rm mis}^{2} \rangle \rightarrow \pi^{2}/3 $ \cite{Marsh:2015xka}.
The total amount of the dark matter will be the sum of these two: $ \Omega_{a} = \Omega_{a}^{\rm mis} + \Omega_{a}^{\rm str} = (1 + \delta_{\rm dec}) \Omega_{a}^{\rm mis}$.
For the case of very light ALP fields, we can take the results from QCD axion simulations ranging from order 1 to $10^{3}$, which still has large uncertainties originating from the ambiguities of the exact spectrum of instantaneous axion emission. 
For heavier ALPs,
cosmic strings are also short-lived, so $ \delta_{\rm dec} \rightarrow 0$ in this limit.
%

\subsubsection{Case I: $f_{a} < T_{\rm reh}$} 
\label{subsubsection:post_1}

In this case, thermal contributions affect the symmetry breaking. Assuming a thermal potential of the form $V_{T} (\Phi) \sim T^2 \vert \Phi \vert^2 $ {with temperature $T$}, the $U(1)$ symmetry is then restored during reheating.
The $U(1)$ breaking phase transition then occurs at the temperature $T_c\simeq f_a$ with the corresponding Hubble parameter $H_c$ given by $H_c=(\pi^2 g_*{(T_{c})}/90)^{1/2} T_c^2/M_P$. There are two cases depending on the time of oscillation:
\begin{itemize}
    \item $ H_{\rm osc} \lesssim H_{c} $

    The ALP coherent oscillation starts after the phase transition, and the ALP DM abundance is given by Eq.~\eqref{eq:pre_1}, with the misalignment angle replaced to the averaged value $ \langle \theta_i^{2} \rangle = \pi^{2}/3 $.

    \item $  H_{\rm osc} \gtrsim H_{c} $

    In this case, the ALP coherent oscillation starts right after the phase transition.
    Hence, the ALP abundance is given by
    \begin{align}
 \left. \frac{\rho_{a}}{s}  \right\vert_{\rm c} 
 & =\frac{ \langle \theta_{i}^{2} \rangle f_{a}^{2} m_{a}^{2}/2}{2\pi^{2} g_{*,s}(T_{c}) T_{c}^{3} / 45 }  
 = \frac{15 m_{a}^{2}}{4 g_{*,s}(T_{c}) f_{a}} \,.
 \label{eq:post_2}
\end{align}

\end{itemize}

\subsubsection{Case II: $T_{\rm reh} < f_a < T_{\rm dS}$} 
\label{subsubsection:post_2}

In this regime, the quantum fluctuations of the ALP field during inflation era exceeds the $U(1)$ breaking scale $f_{a}$ 
\begin{align}
    \delta a \sim \frac{H_{\mathrm{inf}}}{2 \pi } = T_{\mathrm{dS}} > f_{a}.
\end{align}
This leads to a diffused fluctuation of the $\Phi$ field to be centered at zero, effectively restoring the $U(1)$ symmetry~\cite{Kawasaki:2013ae, DiLuzio:2020wdo}. This again leads to a randomly distributed ALP field value, with the average misalignment of $\langle \theta_{\rm mis}^{2} \rangle =  \pi^{2}/3$. 
Once again, we can further specify considering the epoch of oscillation. 

\begin{itemize}
    \item $ H_{\rm osc} \lesssim H_{\rm reh} $: the result is the same as Eq.~\eqref{eq:pre_1}, with $\langle \theta_{\rm mis}^{2} \rangle =  \pi^{2}/3$.

    \item $ H_{\rm osc} \gtrsim H_{\rm reh} $: the result is the same as Eq.~\eqref{eq:pre_2}, with $\langle \theta_{\rm mis}^{2} \rangle =  \pi^{2}/3$.
\end{itemize}


\section{Generic Inflation}
\label{sec:generic_inflation}


As analyzed in the previous section, the scale of inflation heavily affects the abundance of these wormhole-induced ALPs.
In addition, it will also be crucial {whether the radial mode plays the role of the inflaton field or not}.
We first investigate possible constraints for generic inflation cases, with the $U(1)$ scalar being a spectator field in this section.
We then proceed with the case when the $U(1)$ scalar $\Phi$'s radial mode $\rho$ acts as the inflaton with a non-minimal coupling to gravity in Section~\ref{sec:radial_mode_inflation}. 

For pre-inflationary ALP DM production scenarios, one of the main constraints comes from isocurvature perturbations. The null-observation of these perturbations in the CMB leads to a non-trivial bound.\footnote{For the QCD axion, it typically provides an upper bound on the scale of the inflation as $  H_{\rm inf} \lesssim 5.7 \times 10^{8} ( 5 \, \text{neV} / m_{a} )^{0.4175} \, \text{GeV} $ \cite{ParticleDataGroup:2024cfk,Hertzberg:2008wr,Hamann:2009yf}. }
The isocurvature fraction $\beta_{\mathrm{iso}}$ from the ALP in the case of slow-roll inflation and its constraints from Planck observations are given by \cite{Beltran:2006sq,Fairbairn:2014zta}
\begin{equation}
    \begin{aligned}
    \beta_{\rm iso} 
    = 
    \left( 1 + \frac{8 \pi^{3} A_{s} f_{a}^{2} \langle \theta_{i}^{2} \rangle}{ ( \Omega_{a} / \Omega_{\rm CDM})^{2}  H_{\rm inf}^{2}  }\right)^{-1}
    \simeq \left( \frac{\Omega_{a}}{\Omega_{\rm CDM}} \right)^{2} \frac{H_{\rm inf}^{2}}{8 \pi^{3} A_{s} f_{a}^{2} \langle \theta_{i}^{2} \rangle}
    < 0.038
\end{aligned}    \label{eq:isocurvature}
\end{equation}
where $\Omega_{\rm CDM}$ is the ratio of the total dark matter energy density with respect to the critical energy density, $A_{s} \simeq 2.1 \times 10^{-9}$ \cite{Planck:2018jri} is the amplitude of the primordial scalar power spectrum
and we took the constraints on isocurvature from Ref.~\cite{Planck:2018jri}.

\begin{figure}
    \centering
    \includegraphics[width=0.45\textwidth]{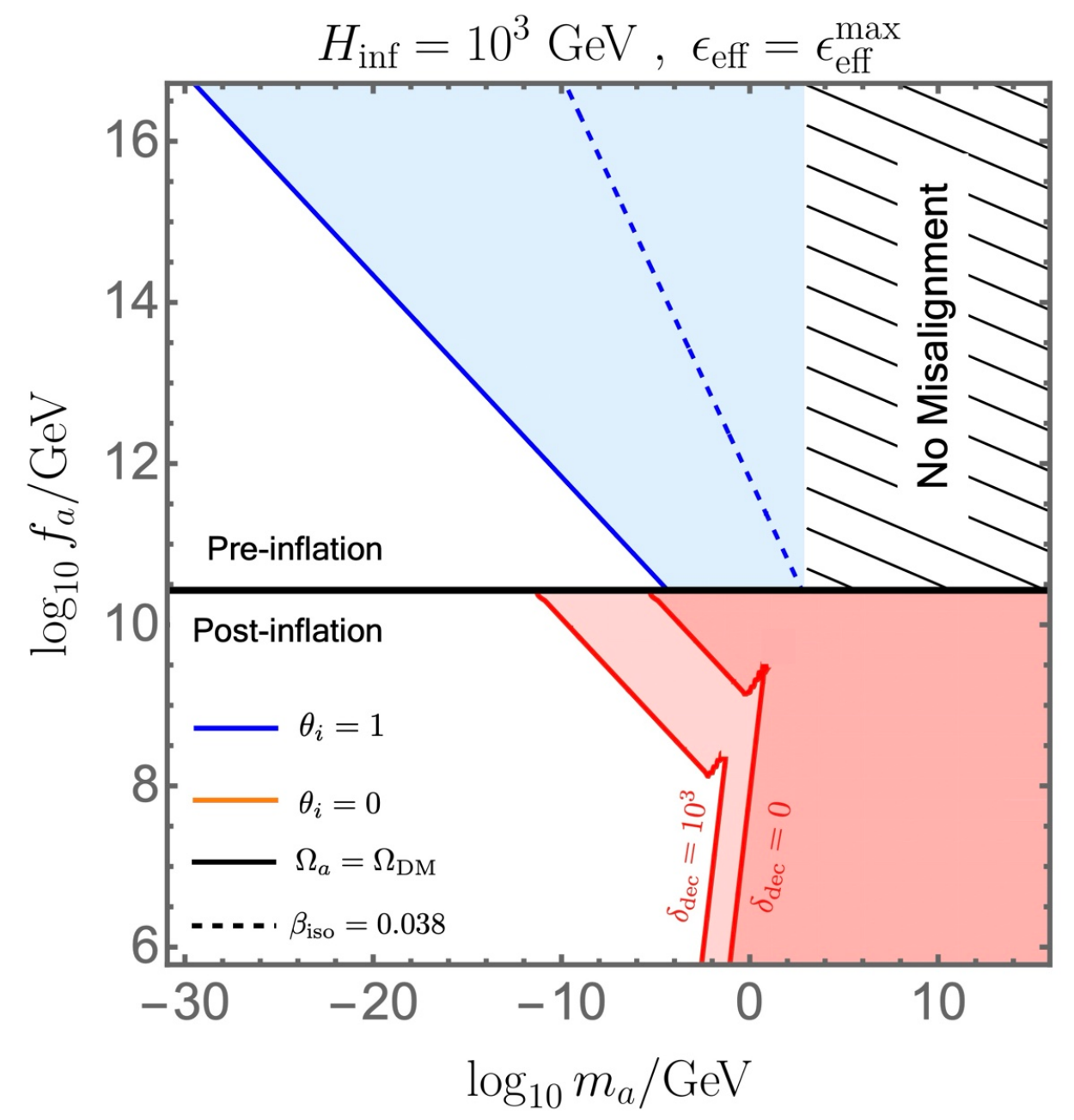}
    \includegraphics[width=0.45\textwidth]{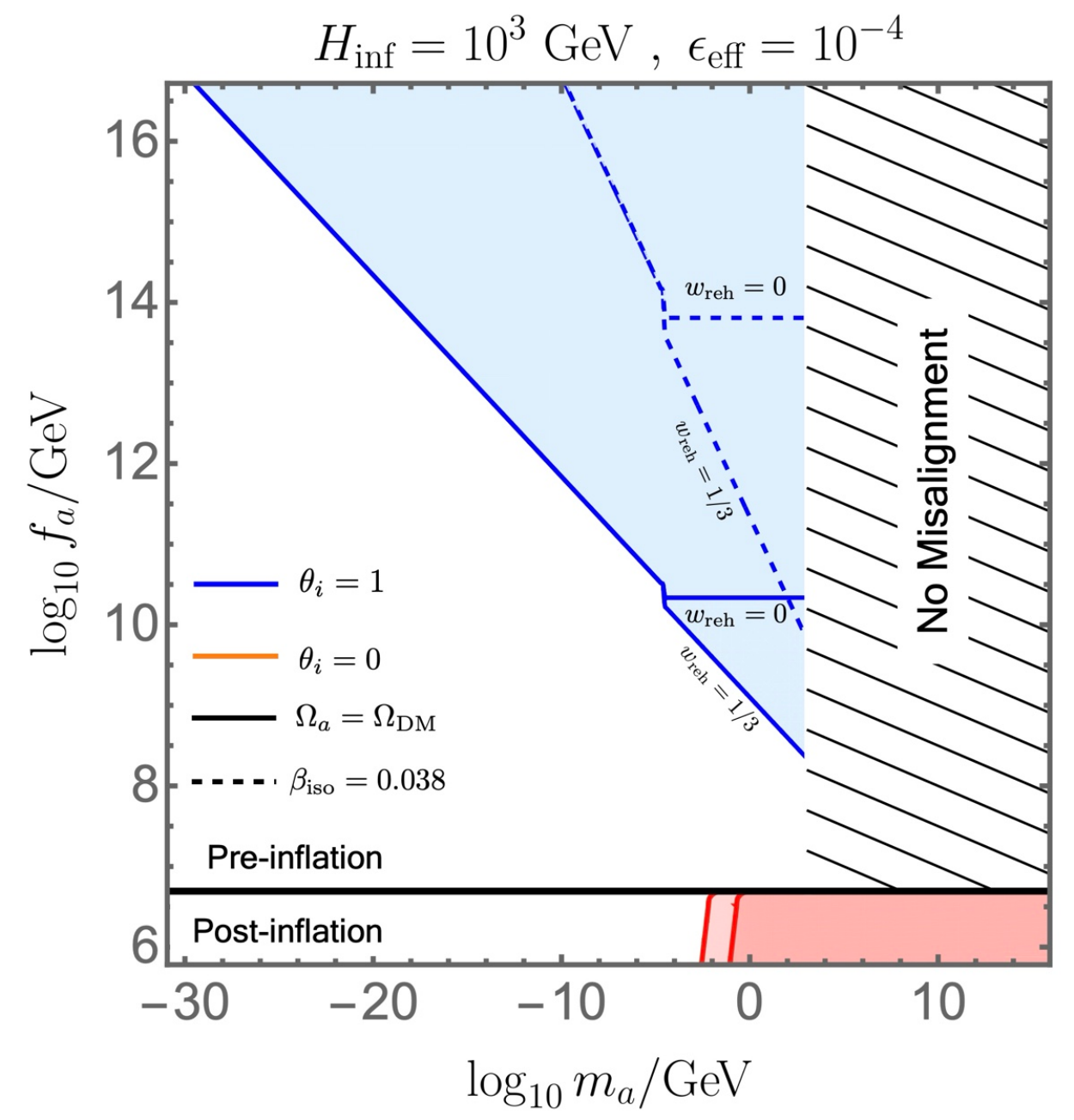}
    \includegraphics[width=0.45\textwidth]{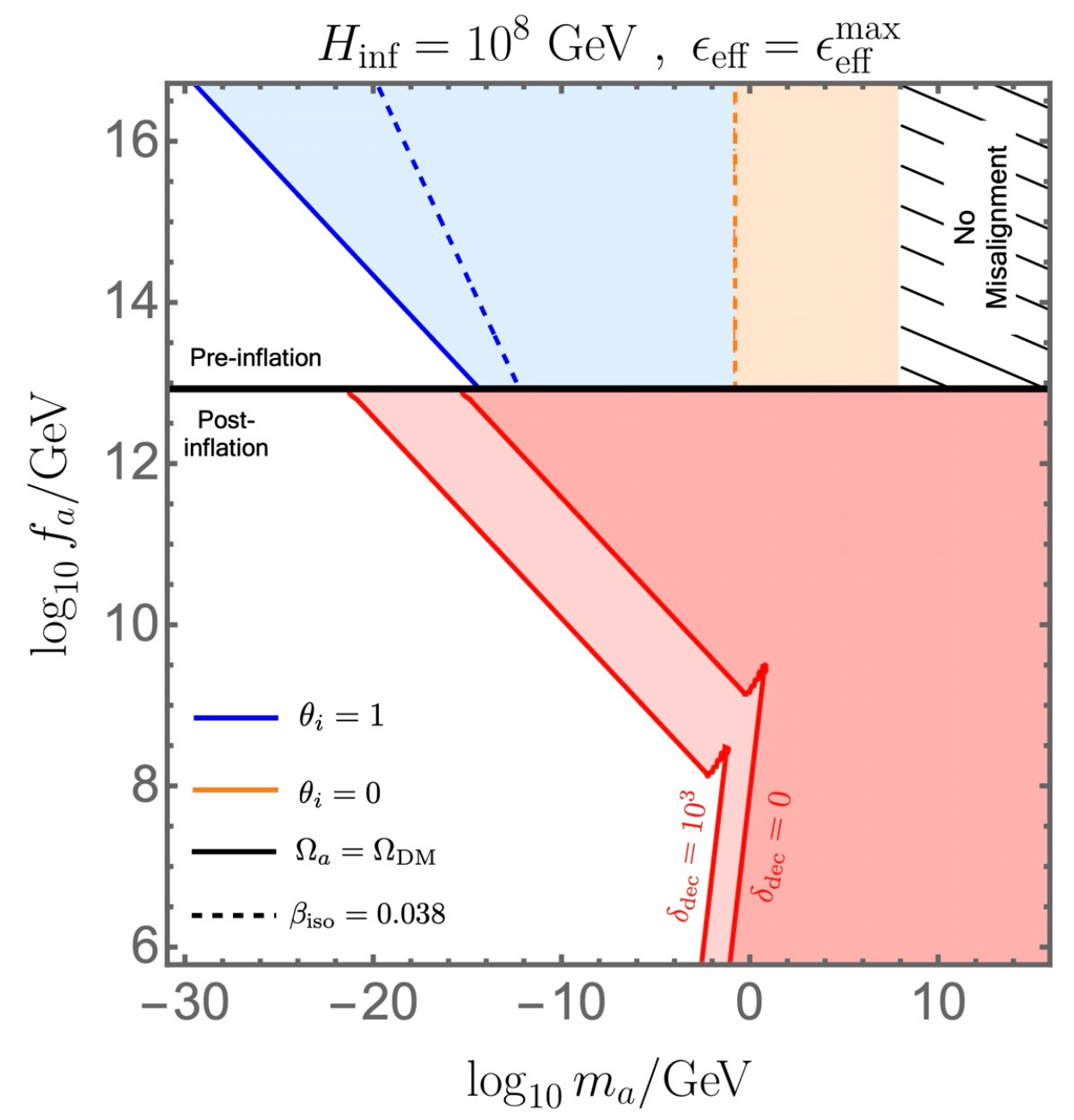}
    \includegraphics[width=0.45\textwidth]{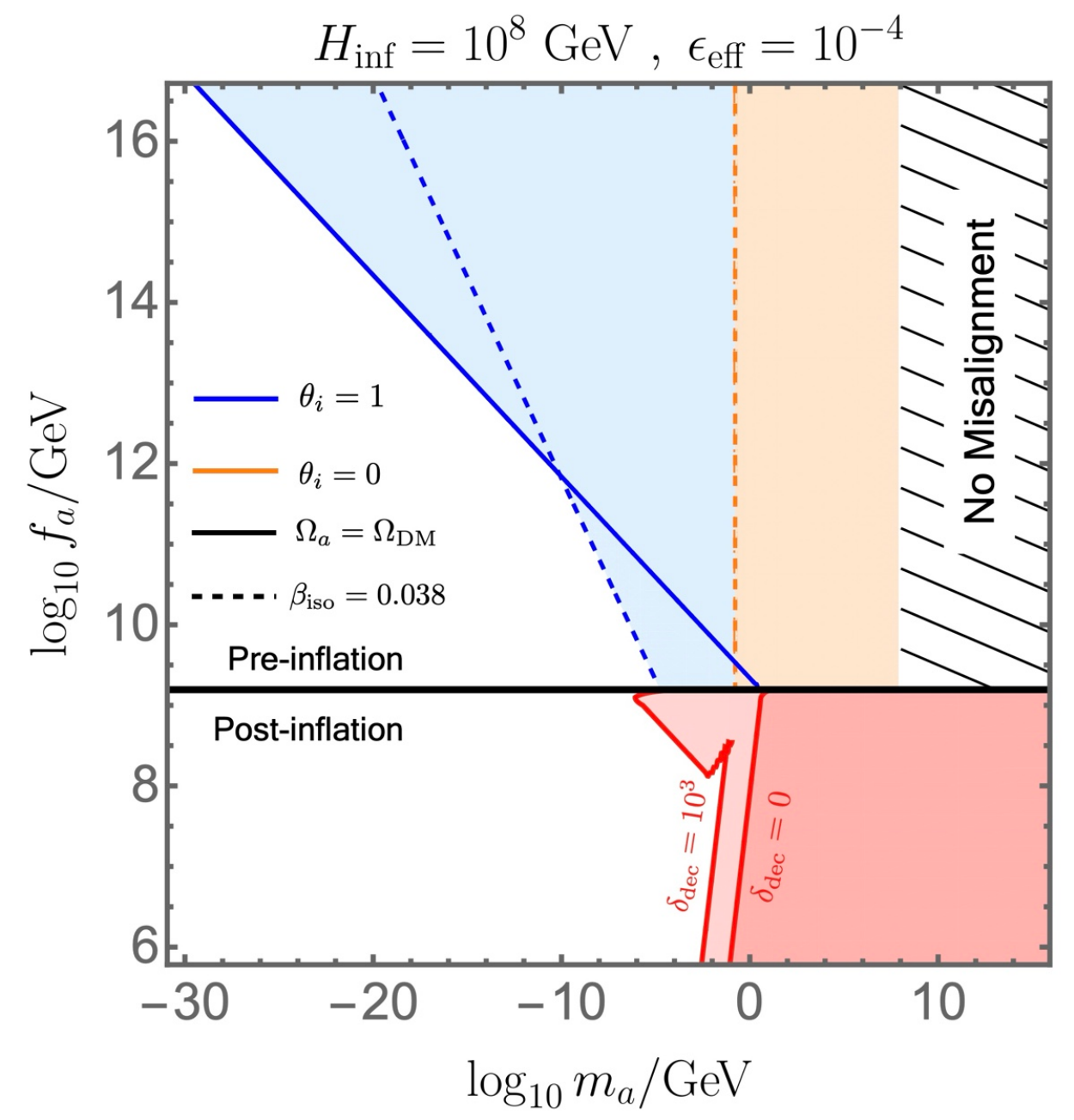}
    \includegraphics[width=0.45\textwidth]{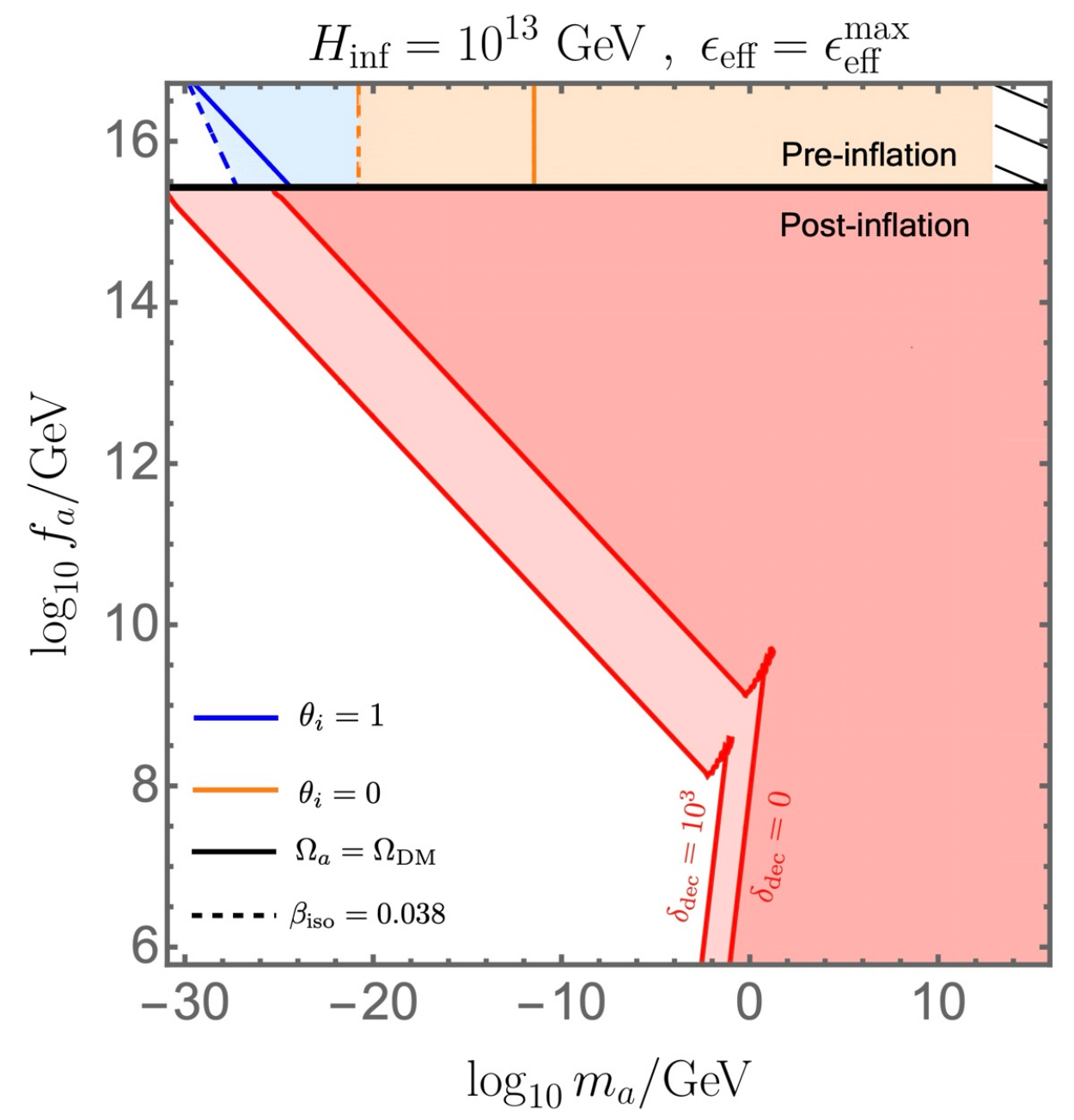}
    \includegraphics[width=0.45\textwidth]{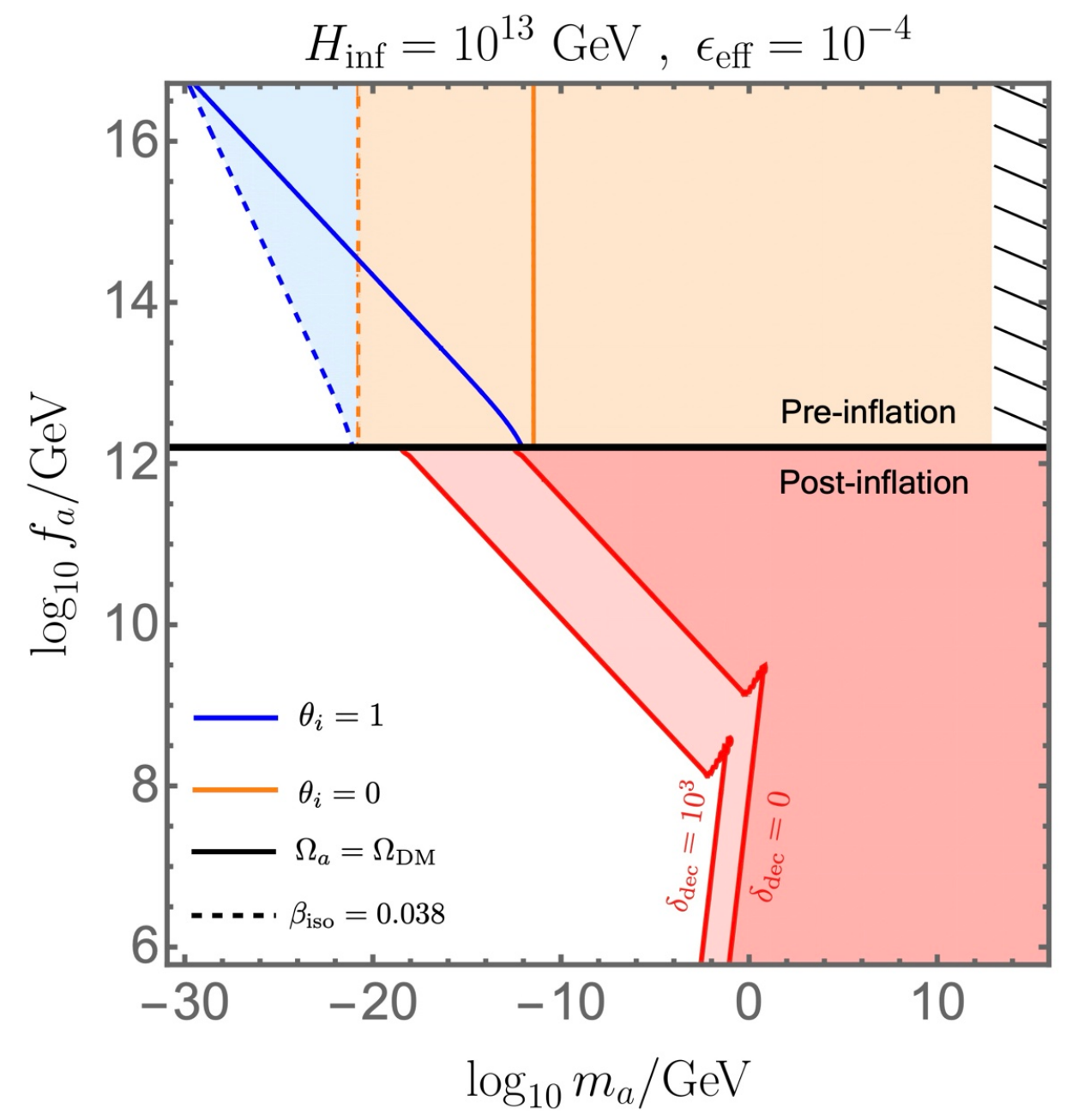}
\caption{
$m_{a}-f_{a}$ contours for DM abundance (solid) and isocurvature constraints (dashed) for $H_{\rm inf} = 10^{3}~{\rm GeV}$ (top), $H_{\rm inf} = 10^{8}~{\rm GeV}$ (middle), $H_{\rm inf} = 10^{13}~{\rm GeV}$ (bottom), with $\epsilon_{\rm eff} = \epsilon_{\rm eff}^{\rm max} $ (left) and $\epsilon_{\rm eff} = 10^{-4}$ (right).
Blue regions represent $\theta_{i} = 1$, and orange regions correspond to $\theta_{i} = 0$, with the latter irrelevant for $H_{\rm inf} = 10^{3}~\rm GeV$ as it resides in the no misalignment region. The $w_{\rm reh} $ dependence is explicitly depicted on the top right figure. }
\label{fig:model_indep}
\end{figure}

\begin{figure*}
    \centering
    \includegraphics[width=0.45\textwidth]{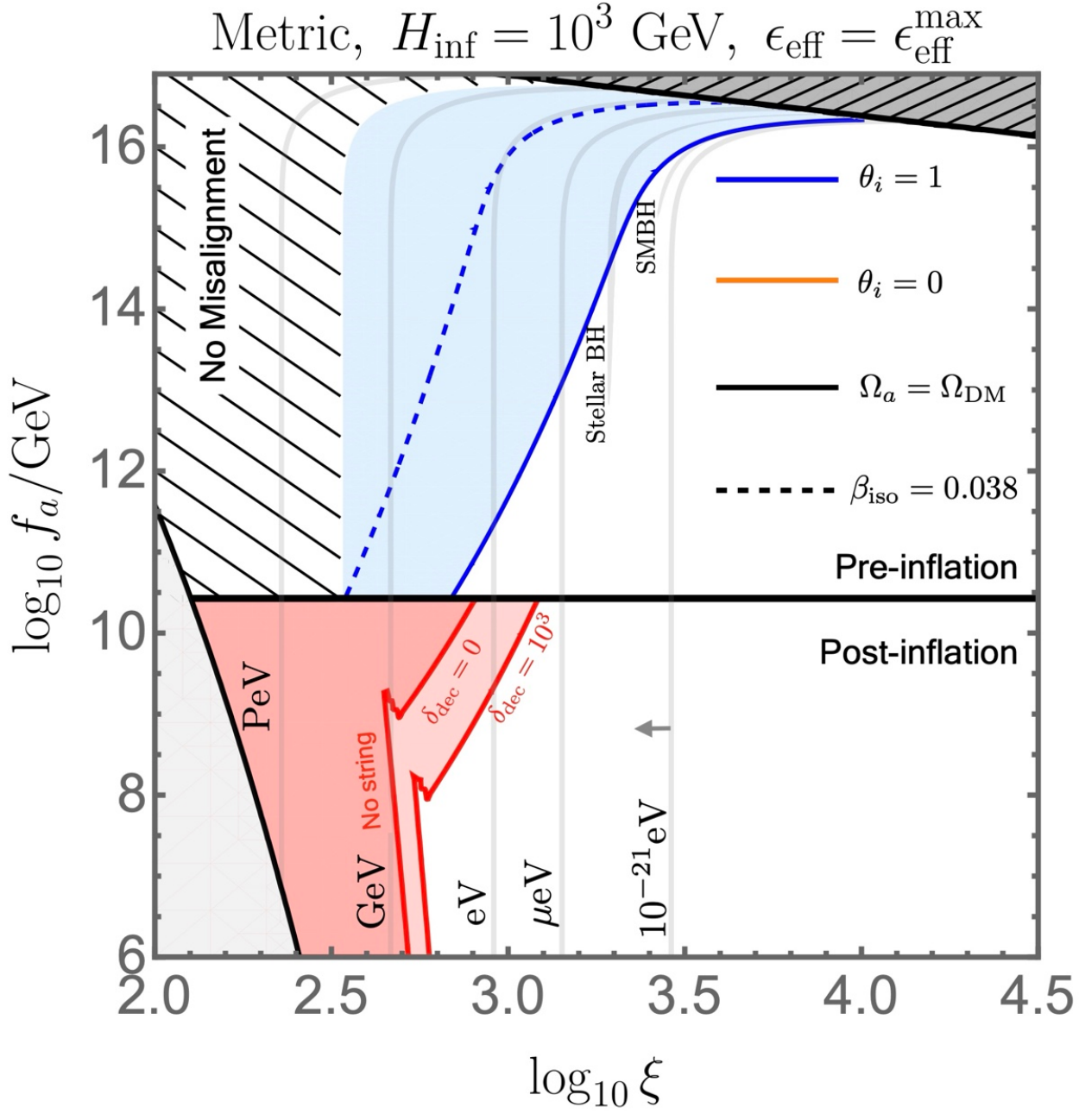}
    \includegraphics[width=0.45\textwidth]{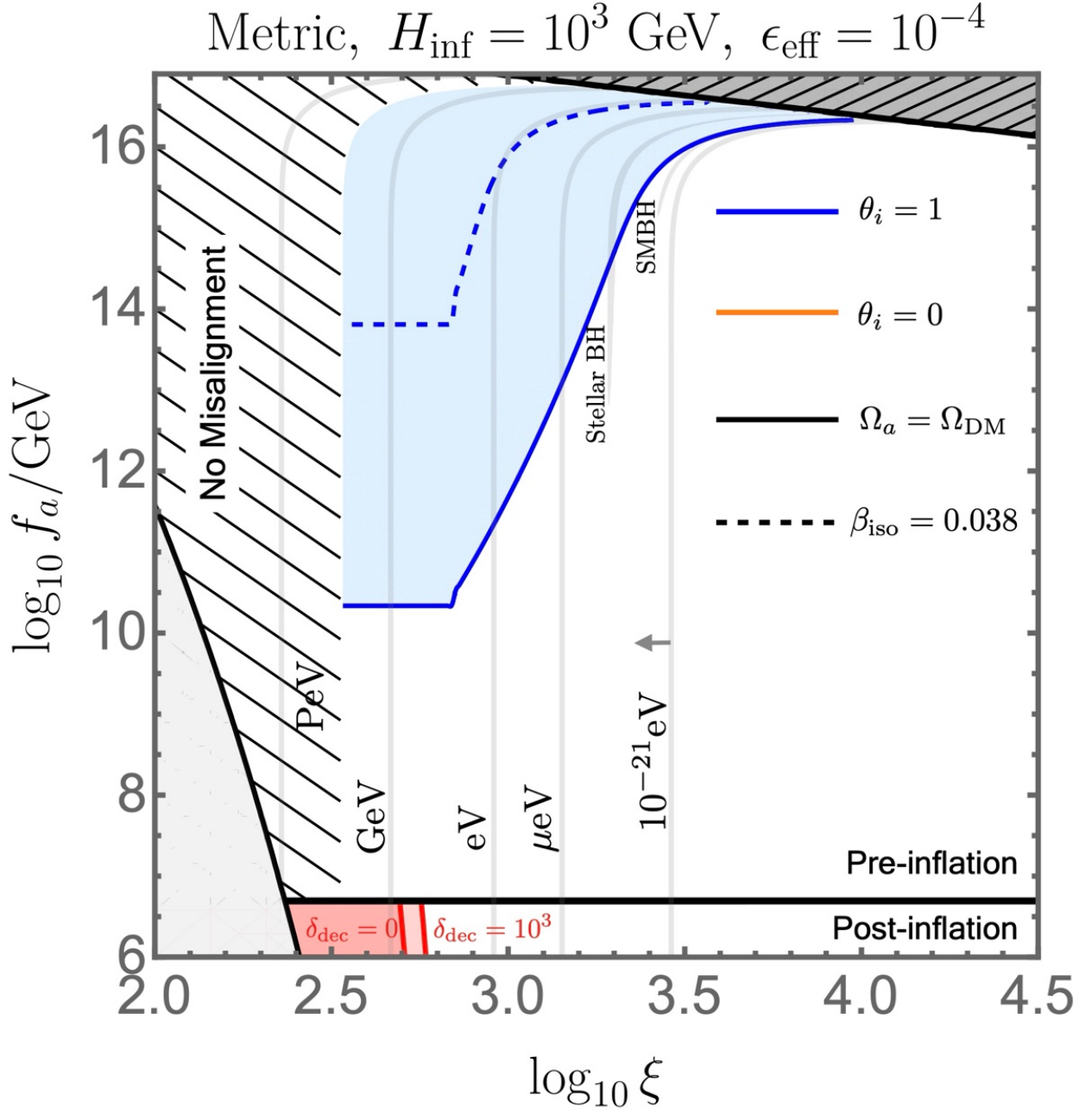}
    \includegraphics[width=0.45\textwidth]{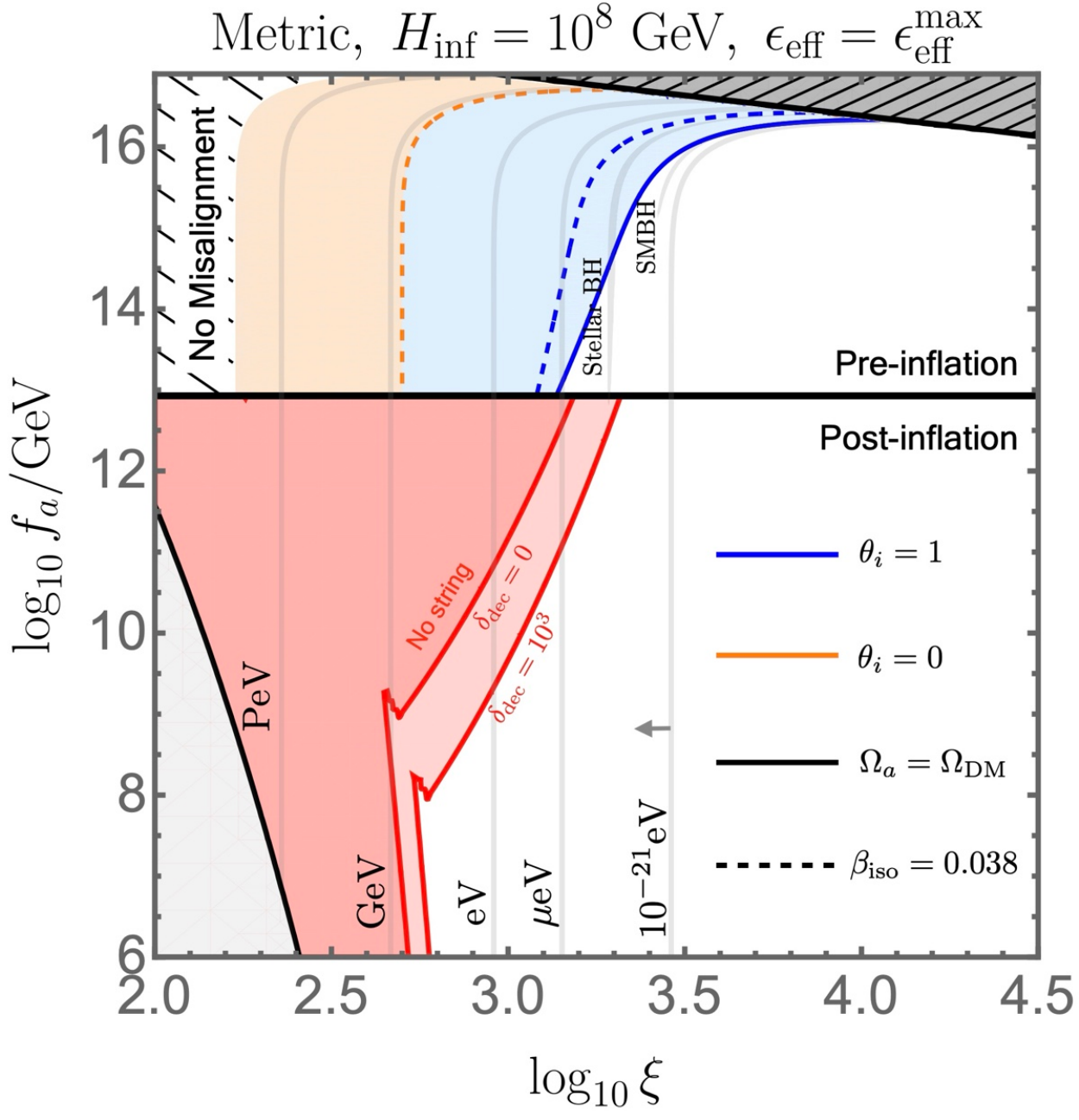}
    \includegraphics[width=0.45\textwidth]{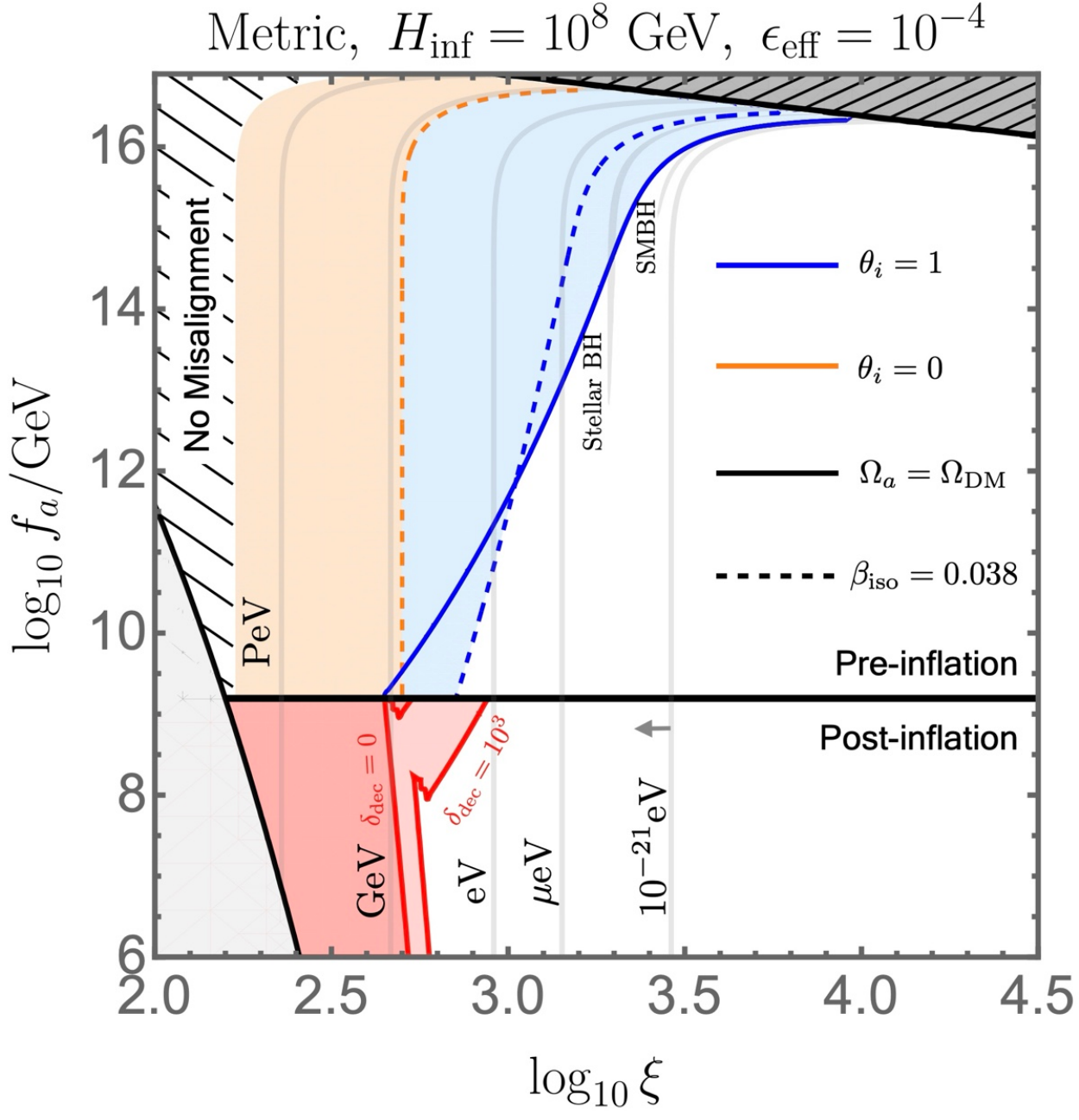}
    \includegraphics[width=0.45\textwidth]{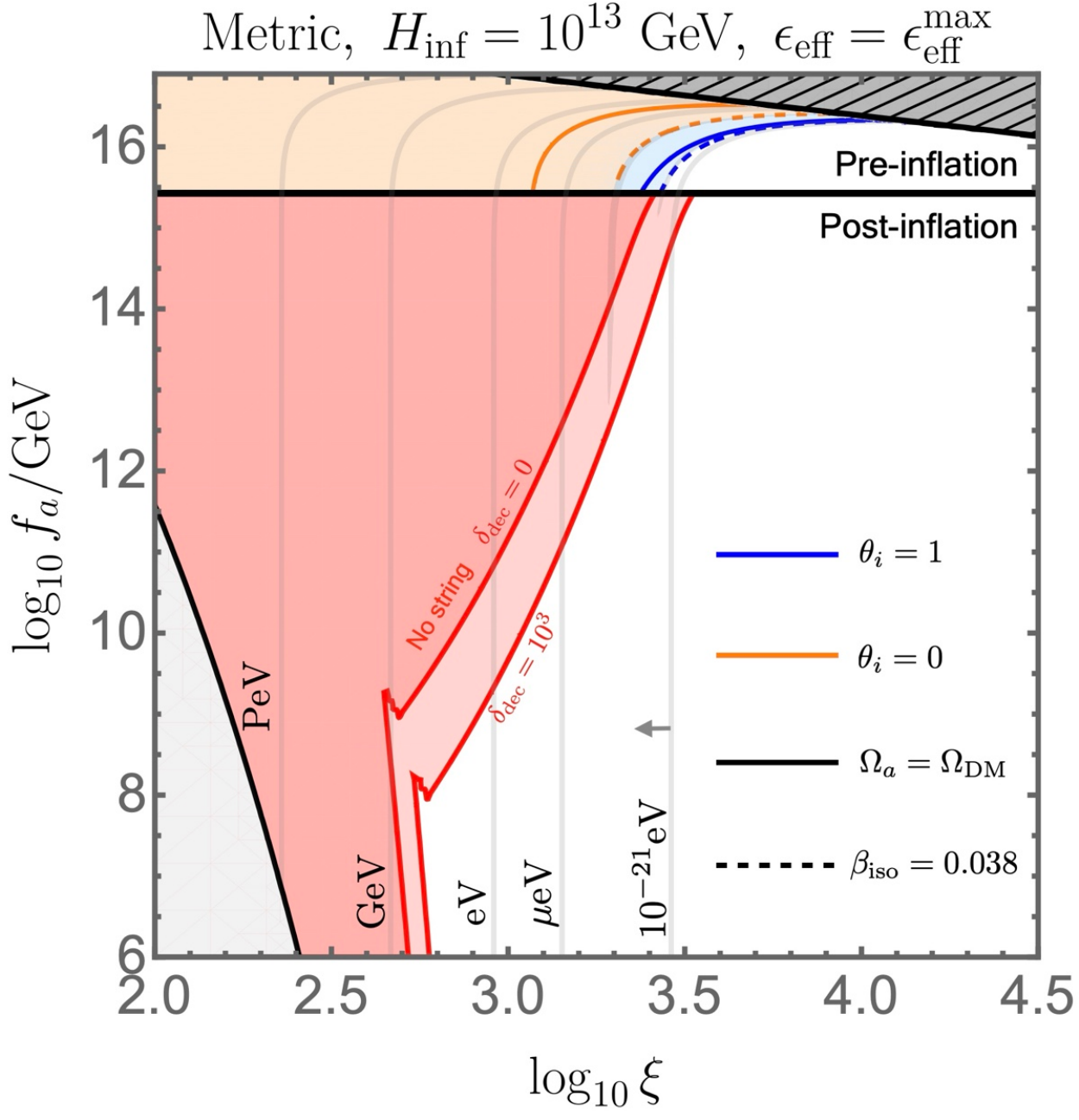}
    \includegraphics[width=0.45\textwidth]{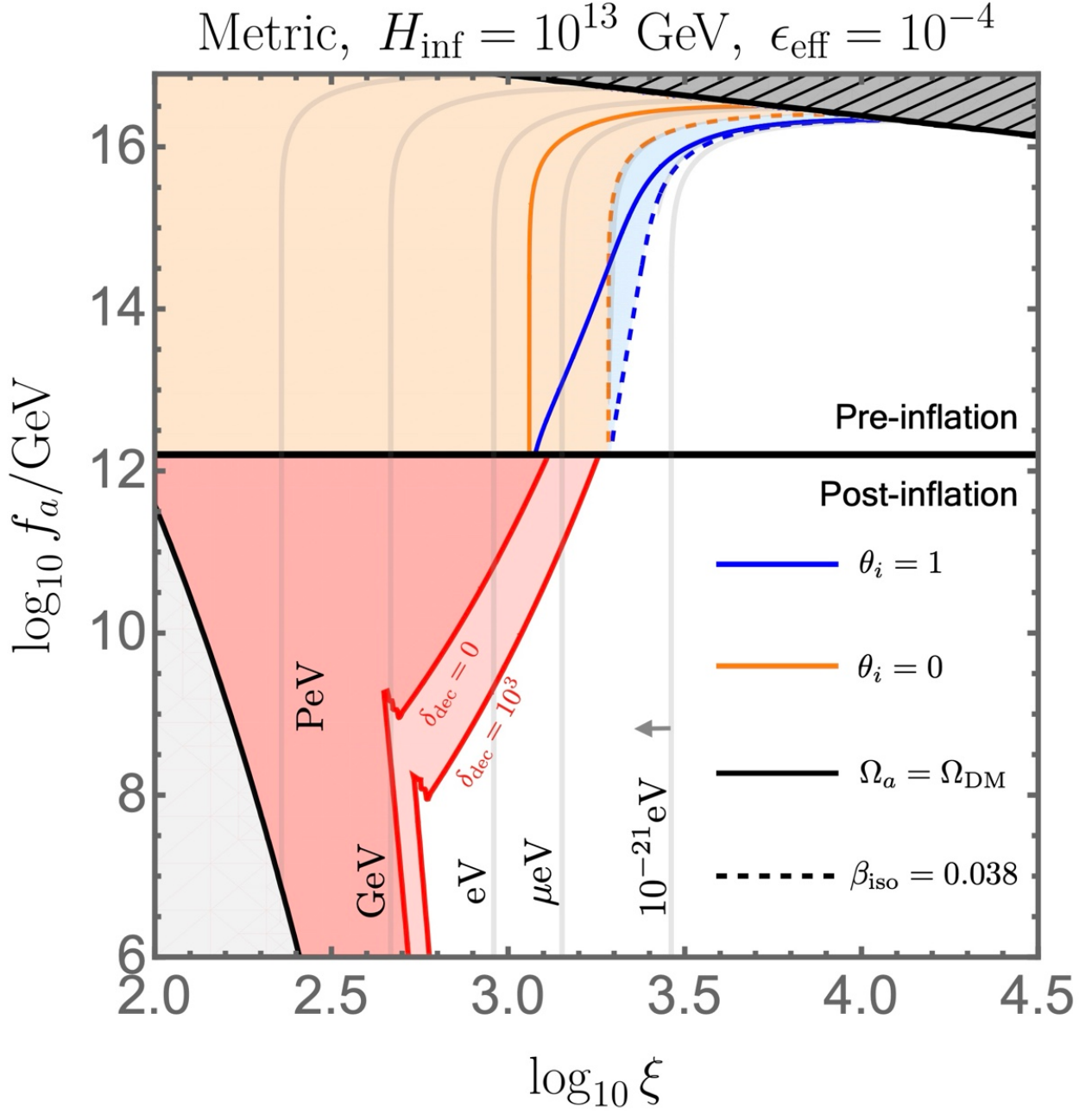}

    \caption{Constraints for generic inflation in the metric formalism, with $w_{\rm reh } = 0 $, $\lambda_{\Phi} = 10^{-2} $, and mass contours (light gray). Colored regions are constrained, with solid lines giving $\Omega_{a} = \Omega_{\rm DM} $ and dashed lines $\beta_{\rm iso} = 0.038$.
    Different color schemes for pre-inflation represent different $\theta_{i}$ cases.
    The post-inflationary cases contain both misalignment and cosmic string contributions with $\Omega_{a}^{\rm tot} = ( 1 + \delta_{\rm dec} ) \Omega_{a}^{\rm mis}
    $.
    The gray-shaded regions in the bottom-left corners correspond to the regions of $m_\rho < m_a$.
}

    \label{fig:generic_metric_1}
\end{figure*}

\begin{figure*}
    \centering

    \includegraphics[width=0.45\textwidth]{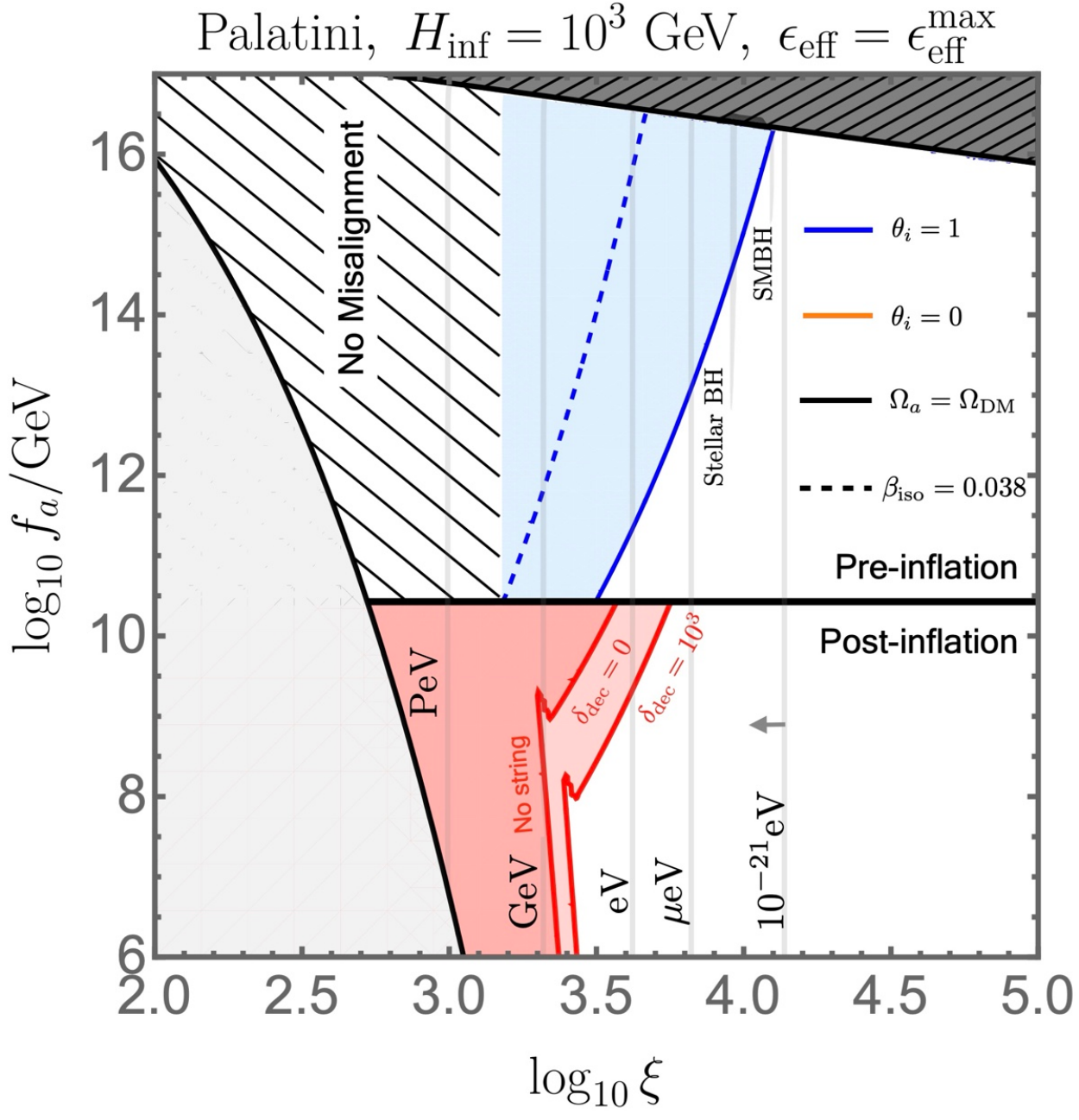}
    \includegraphics[width=0.45\textwidth]{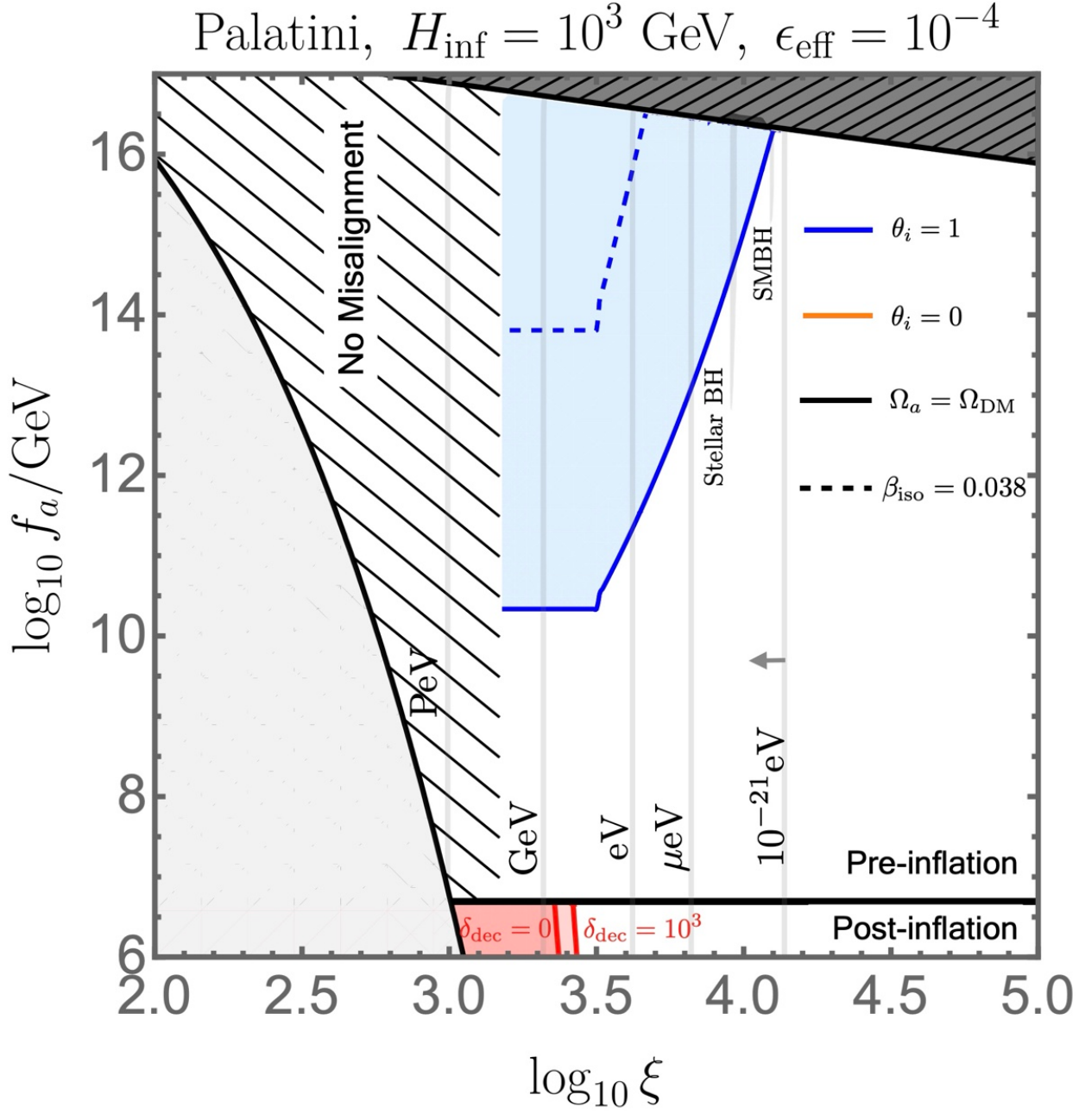}
    \includegraphics[width=0.45\textwidth]{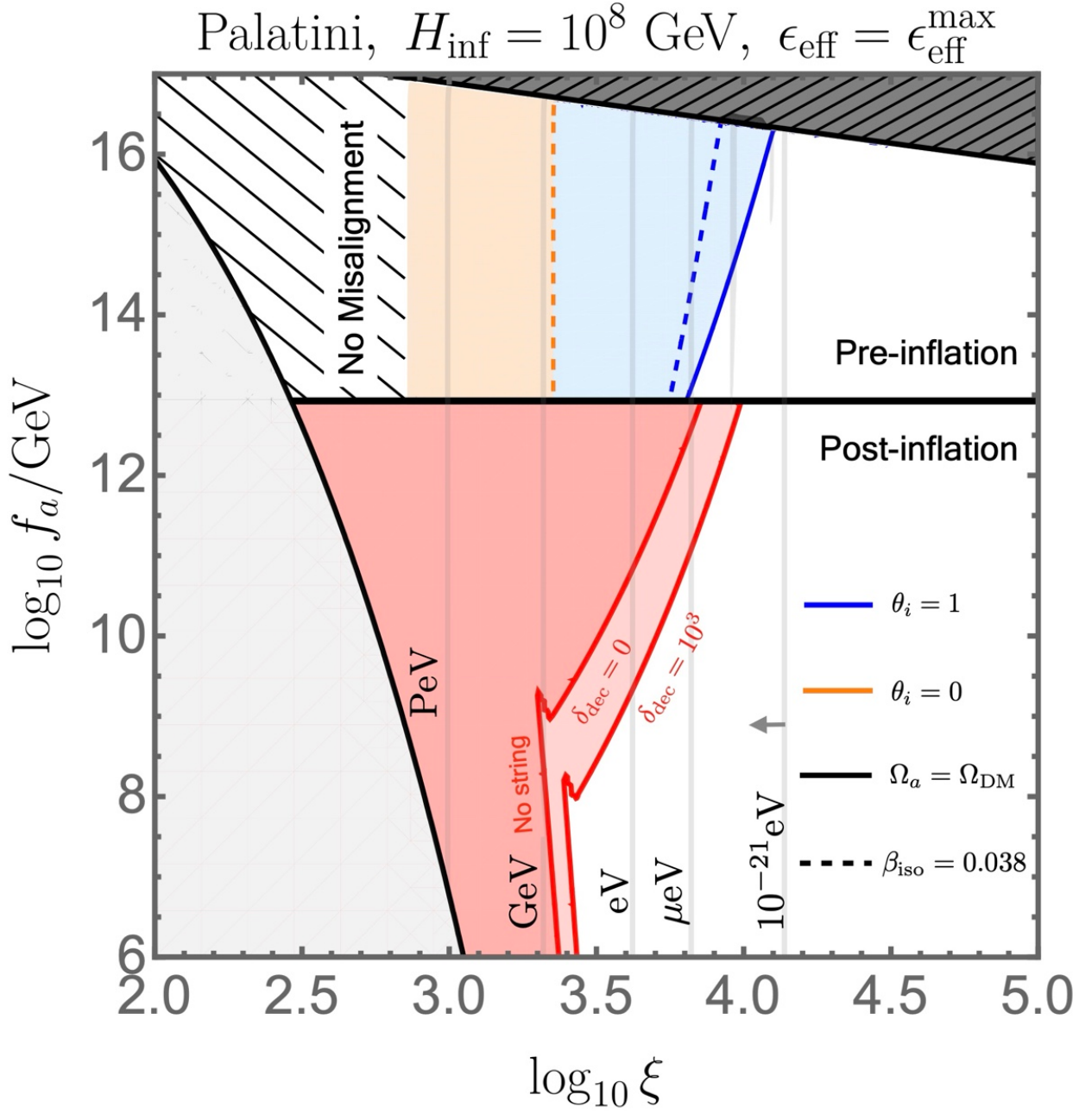}
    \includegraphics[width=0.45\textwidth]{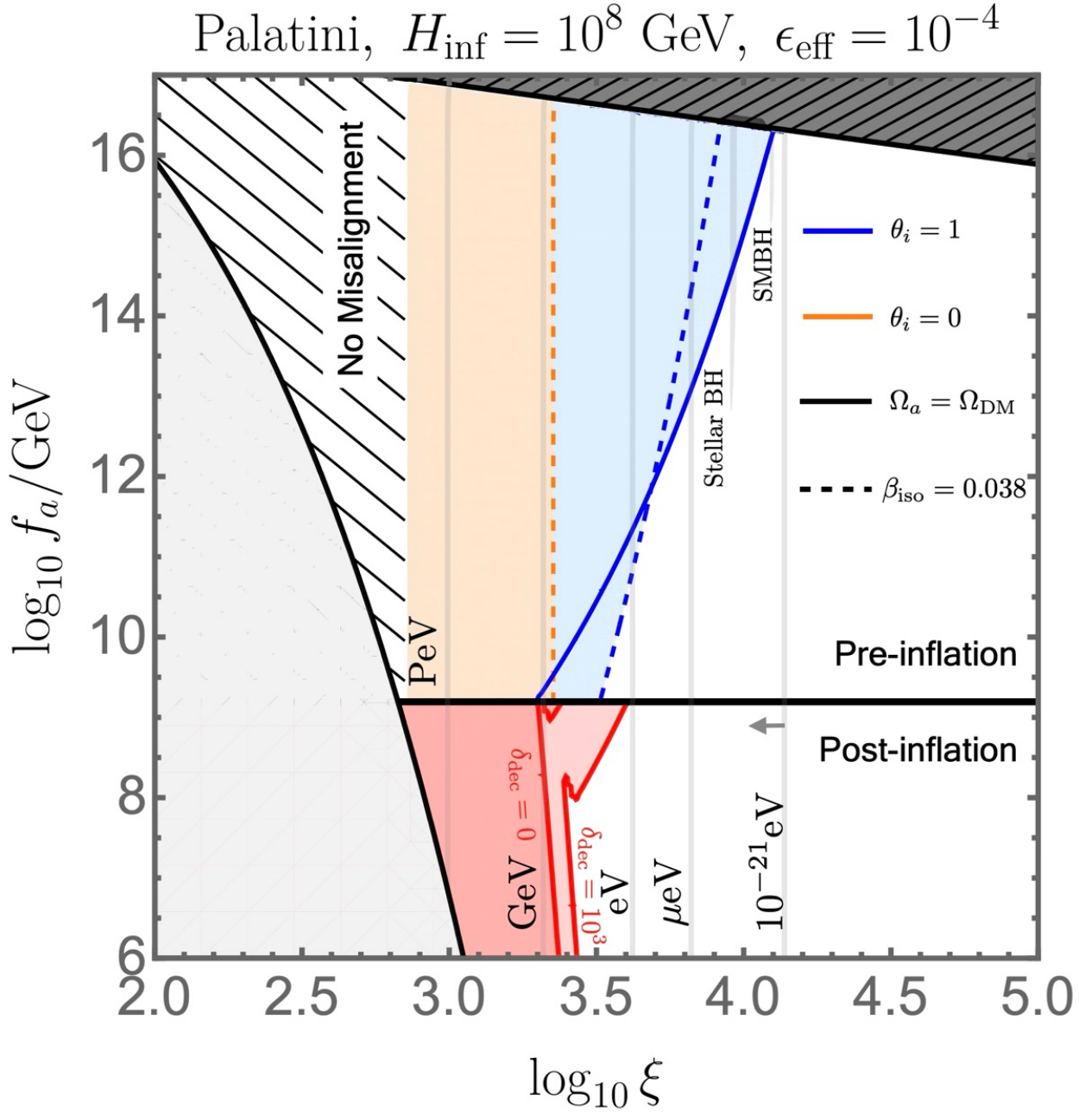}

    \includegraphics[width=0.45\textwidth]{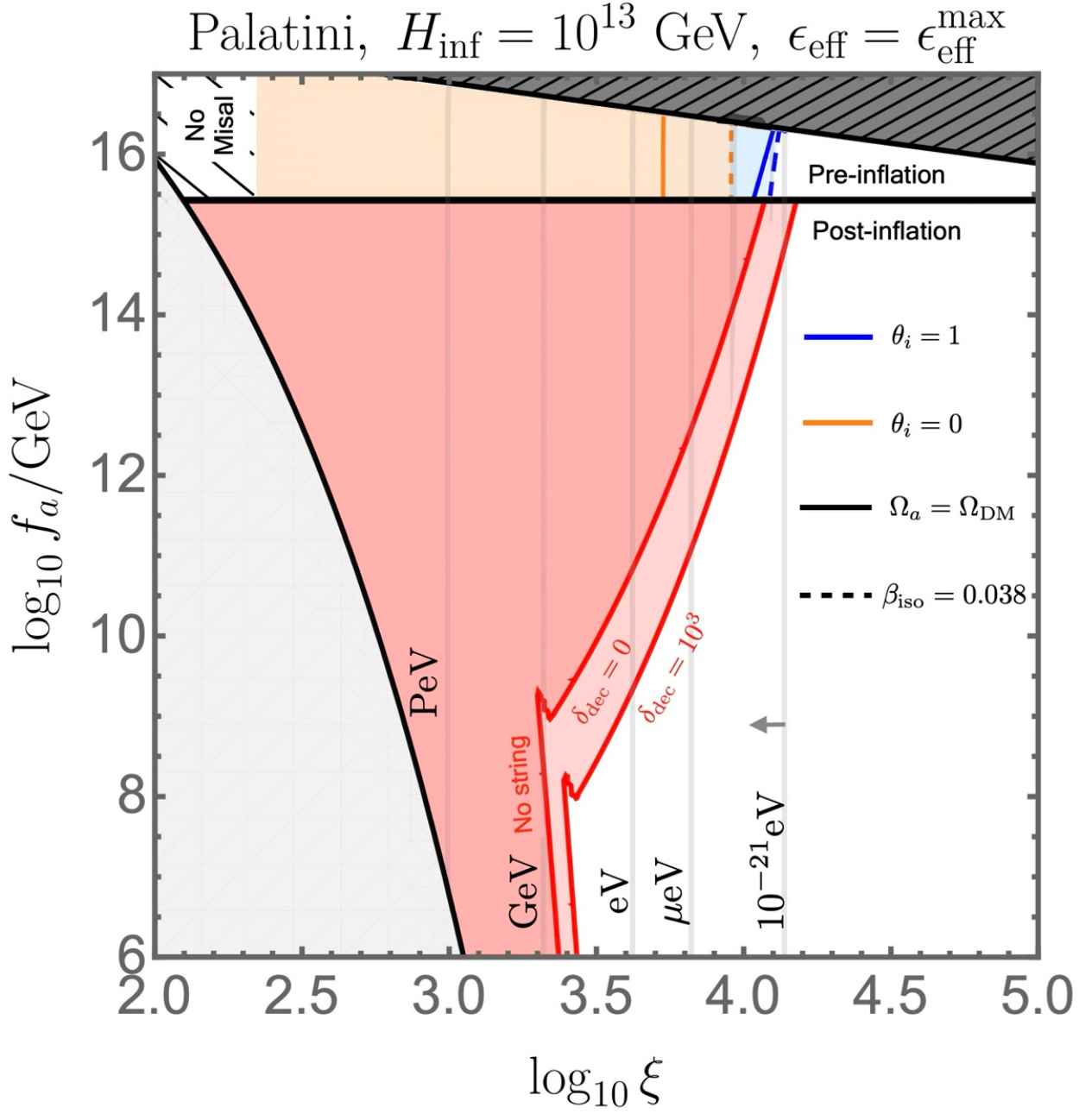}
    \includegraphics[width=0.45\textwidth]{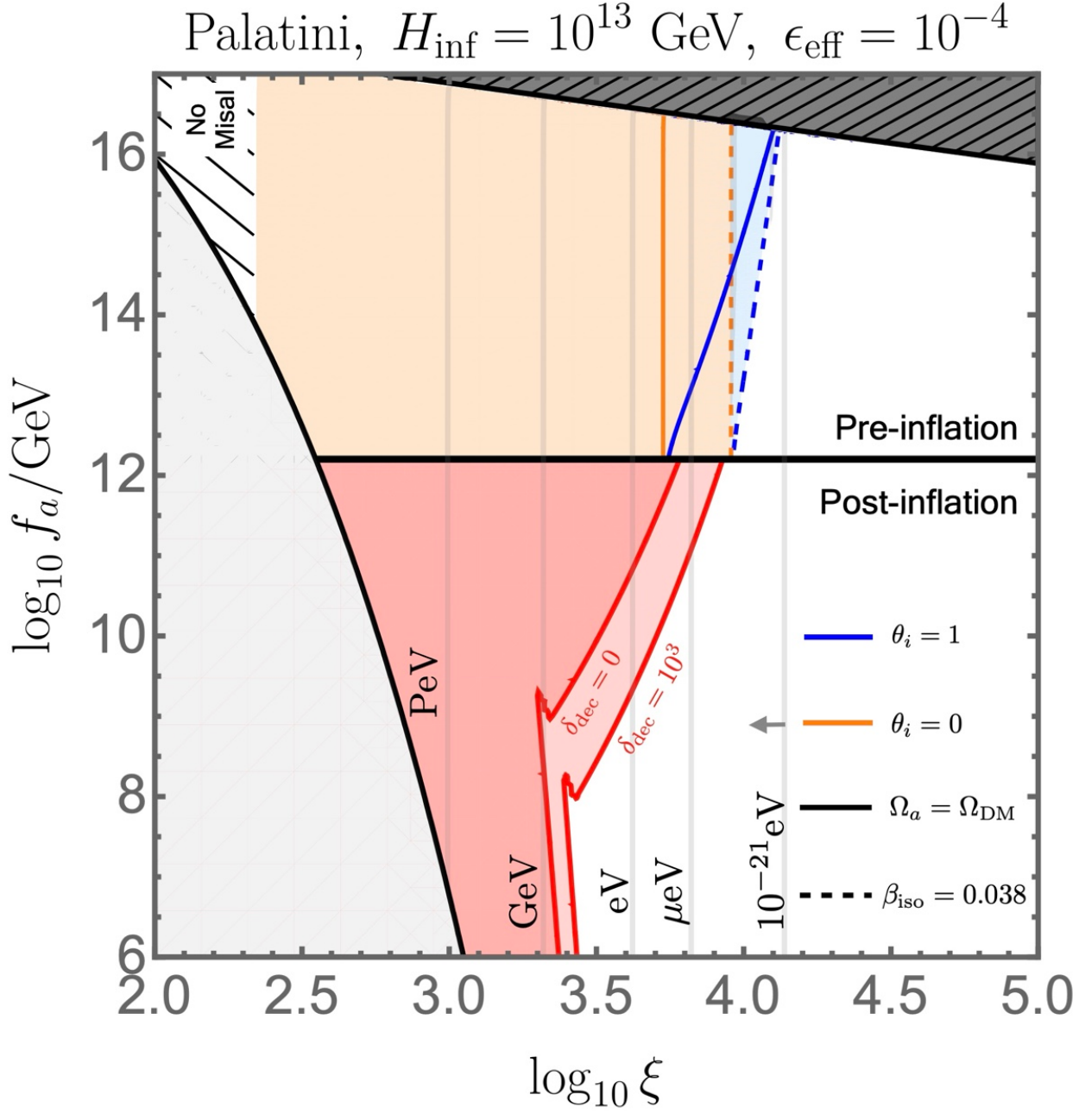}

    \caption{ Same plot scheme as in Figure~\ref{fig:generic_metric_1}, but in the Palatini formalism.
    \label{fig:generic_Pala_1}
}\end{figure*}

Combining the DM abundance formulae Eqs.~\eqref{eq:pre_1}, \eqref{eq:pre_2} and Eq.~\eqref{eq:isocurvature}, we have bounds on the $ H_{\rm inf} $ in terms of $m_{a}$ as
\begin{align}
   H_{\rm inf} \lesssim 10^{13}  \left( \frac{\Omega_{a}}{\Omega_{\rm CDM}} \right)^{-\frac{1}{2}} 
    \left( \frac{\Omega_{\rm CDM} h^{2}}{0.12} \right)^{\frac{1}{2}} 
   \left( \frac{A_{s}}{2.1 \times 10^{-9}} \right)^{\frac{1}{2}} \left( \frac{\beta_{\rm iso}}{0.038} \right)^{\frac{1}{2}} \left( \frac{m_{a}}{10^{-21} \, \text{eV}} \right)^{-\frac{1}{4}}  \, \text{GeV} \, ,
   \label{eqn:Hboundisocurvature}
\end{align} 
for $H_{\rm osc}<H_{\rm reh}$ and
\begin{align}
    H_{\rm inf} \lesssim  1.5 \cdot 10^{4} \left( \frac{\epsilon_{\rm eff}}{1} \right)^{-2/5} \left( \frac{\Omega_{a}}{\Omega_{\rm CDM}} \right)^{ -\frac{2}{5}} 
   \left( \frac{\Omega_{\rm CDM} h^{2}}{0.12} \right)^{\frac{2}{5}}
   \left( \frac{A_{s}}{2.1 \times 10^{-9}} \right)^{\frac{2}{5}} \left( \frac{\beta_{\rm iso}}{0.038} \right)^{\frac{2}{5}}  \, \text{GeV} \, . \label{eqn:Hboundisocurvature_2}
\end{align}
for $H_{\rm reh}<H_{\rm osc}<H_{\rm inf}$
where we took $w_{\rm reh} = 0$ for simplicity in Eq.~\eqref{eqn:Hboundisocurvature_2}.
Note that this is in contrast with the conventional QCD axion bound. By allowing $ m_{a} $ and $ f_{a} $ to be independent and $ m_{a} $ to be temperature independent, larger values of the Hubble parameter are still compatible.
One thing to mention is that the isocurvature bound only holds when $m_{a} < H_{\rm inf}$, because for larger mass, the quantum fluctuation is suppressed during inflation.
Now that we have all the necessary formulas for the ALP dark matter abundance and constraints, let us turn to a discussion of the results. We first show the constraints in the $(m_a, f_a)$ planes in Figure~\ref{fig:model_indep}.
This is a generic result and independent of the origin of the ALP mass. We then present the results for the wormhole-induced ALP cases in the $(\xi, f_a)$ planes by using the mass relation Eq.~(\ref{eq:ALPmasswh}), 
in Figure~\ref{fig:generic_metric_1} for metric cases and Figure~\ref{fig:generic_Pala_1} for Palatini cases, respectively.
In all cases, we depict the constrained regions taking large $\theta_{i} = 1$ and small $\theta_{i} = 0$ 
for different $H_{\rm inf} = (10^{3}, \, 10^{8}, \, 10^{13}) \, {\rm GeV}$.
Each figure for $H_{\rm inf}$ value considers two cases for $\epsilon_{\rm eff}$, $\epsilon_{\rm eff}=\epsilon_{\rm eff}^{\rm max}$ and $\epsilon_{\rm eff}=10^{-4}$. 
We start with the model-independent ALP cases in Figure~\ref{fig:model_indep}.
For $H_{\rm inf} = (10^{3}\, , 10^{8})~\rm GeV$, the pre-inflationary scenario exhibits 100\% DM abundance for large misalignment values $\theta_{i} = 1$. However, for the case $\theta_{i} = 0$, the misalignment production only with quantum fluctuation is insufficient to explain the entirety of dark matter.
Additional constraints enter from isocurvature perturbations, which further restricts all dark matter for $m_{a} \gtrsim 0.1\rm eV$ with $\theta_{i} = 1$, as shown in the $H_{\rm inf}  = 10^{8}~\rm GeV$, $\epsilon_{\rm eff} =10^{-4} $ case.
For $H_{\rm inf}=10^{13}\,{\rm GeV}$, isocurvature fluctuation constraints dominate over DM abundance curves for both $\theta_{i} = 1$ and $\theta_{i} = 0$ cases, removing the possibility for these ALPs to explain our DM abundance.
The constraints also depend on the equation of state during the reheating $w_{\rm reh }$ for ALPs starting to oscillate during this period, as depicted in the $H_{\rm inf}  = 10^{3}~\rm GeV$, $\epsilon_{\rm eff} =10^{-4} $ case.
Post-inflationary scenarios can also explain all dark matter for masses $m_{a} \lesssim \rm GeV$, with the exact range highly depending on $H_{ \rm inf} $ and $\epsilon_{\rm eff}$ values.

Using Eq.~(\ref{eq:ALPmasswh}) we 
can translate the results in Figure~\ref{fig:model_indep} into the ones 
for the wormhole-induced ALP cases, 
which are shown in the $(\xi, f_a)$ planes in Figure~\ref{fig:generic_metric_1} for metric cases and Figure~\ref{fig:generic_Pala_1} for Palatini cases, respectively.
The gray-shaded regions in the bottom-left corners correspond to the regions of $m_\rho < m_a$ for $\lambda_\Phi=10^{-2}$.
For both formalisms in the pre-inflationary scenario, the dependence in the initial misalignment $\langle \theta_{\mathrm{mis}}^2 \rangle $ implies that
having smaller $\langle \theta_{\mathrm{mis}}^2 \rangle $s
shifts overall constraints to heavier masses, consequently smaller $\xi$ values.
Due to the different $\langle \theta_{\mathrm{mis}}^2 \rangle $ dependence for the ALP abundance and the isocurvature constraints, the most stringent constraint also switches for larger values of $H_{\rm inf} $, where in this case the ALPs induced by wormholes cannot consist the entirety of dark matter.
Therefore, the shaded regions for $\theta_{i} = 1 $ in Figure~\ref{fig:generic_metric_1} and Figure~\ref{fig:generic_Pala_1} could be allowed if one considers a smaller $\theta_{i}$ value, which will shift both DM and isocurvature constraints to smaller $\xi$ values at different rates.
Correspondingly, the $\beta_{\rm iso}$ constraint for $\theta_{i}=0$ provides an irreducible lower bound on the $\xi$ value for allowed ALPs.
For post-inflationary scenarios the overall behavior of the constraints are similar between the two formalisms, with a larger abundance from cosmic string constraining a larger range in the parameter space.
For the metric formalism, the nontrivial $\xi - f_{a} $ dependence near the GS limit also translates over to the constraints. 
The region with $\xi > M_P^2 / f_a^2$ is excluded and corresponds to the hatched gray area in the top-right of the figures.

The different dependencies on the parameters lead to features in the constraints. Starting with the pre-inflationary scenario, we notice a $\xi$ independent region for the $\Omega_{a} = \Omega_{\rm DM} $ contour. This is when the ALP starts to oscillate during the reheating era, with our figures taking $w_{\rm reh} = 0 $. This removes the $m_{a}$ dependence in Eq.~(\ref{eq:pre_2}), ultimately giving a flat plateau in the $\xi- f_{a}$ plane. This can be seen in the top right panels of the Figures~\ref{fig:generic_metric_1} and \ref{fig:generic_Pala_1} in both solid and dashed blue curves.
Focusing on the small $\theta_{i} = 0$ case, we also notice a $f_{a} $ independence in the DM abundance and the isocurvature constraints for $ H_{\rm inf} = (10^{8}, \, 10^{13})~\mathrm{GeV} $ values {apart from the GS limit}. In this regime, the quantum fluctuations dominate the misalignment angle $\langle \theta_{i}^2 \rangle \simeq (H_{\rm inf}  / (2\pi f_{a}) )^{2} $, now canceling the $f_{a}$ dependence in the constraints.
The post-inflationary DM abundance contours also exhibit different $\xi-f_{a}$ dependencies. Most noticeably there is a kink, which occurs as the ALP abundance from Eq.~(\ref{eq:post_2}) becomes a more stringent bound than $H_{\rm osc} > H_{c}$, therefore gives a gap between the constraints coming from $H_{\rm osc} > H_{c}$ cases, and $H_{\rm osc} < H_{c}$ cases. Around this kink there is also a change in the overall proportionality, with the former case the DM constraints have a $f_{a} \propto m_{a}^2 $ dependence, and the latter having a $f_{a} \propto m_{a}^{-1/4} $ dependence.

\section{Radial Mode Inflation}
\label{sec:radial_mode_inflation}

A scalar field with a large non-minimal coupling is widely considered in the context of inflation which gives a consistent fit to CMB observations.
Given that the radial mode $\rho$ has a non-minimal coupling, it could also take the role of the inflaton, {and the consideration of the isocurvature bound significantly differs from  generic inflation cases \cite{Fairbairn:2014zta}}.
To be consistent with the power spectrum at CMB, $A_{s} \simeq 2.1 \times 10^{-9}$ \cite{Planck:2018jri} with an inflation period about 60 e-folds, the parameters $\xi$ and $\lambda_\Phi$ should satisfy~\cite{Bezrukov:2007ep, Bauer:2008zj, Cheong:2021kyc}:
\begin{align}
    \xi \simeq \begin{dcases}
    4.9 \times 10^{4} \sqrt{\lambda_\Phi} & (\text{metric}) \\
    1.4 \times 10^{10} \lambda_\Phi & (\text{Palatini}).
    \end{dcases}
    \label{eq:xiinf}
\end{align}
For large $\xi$, the Jordan frame field value when the perturbations of the CMB pivot scale leave the horizon and the scale of the inflation $H_{\rm inf}$ are given by 
\begin{align}
        \rho_{*} \simeq 
        \begin{dcases}
          \left(\frac{4N_{e}}{3\xi}\right)^{1/2} M_{P} & (\text{metric}) \\  
          2\sqrt{2 N_{e}} M_{P} & (\text{Palatini})
        \end{dcases} ,
\label{eq:field_values}
    \end{align}
and
\begin{align}
    H_{\rm inf} \simeq \begin{dcases}
        \frac{\sqrt{6A_{s}} \pi M_{P}}{N_{e}} \simeq 1.4 \times 10^{13} \, {\rm GeV} & (\text{metric}) \\
        \frac{\sqrt{A_{s}} \pi M_{P}}{N_{e} \sqrt{\xi}} \simeq \frac{ 5.9 \times 10^{12} }{ \sqrt{\xi} } \, {\rm GeV} & (\text{Palatini}) 
    \end{dcases}
    \label{eq:radial_Hinf}
\end{align}
where $N_{e}$ is the e-folding number during the inflation and we took $ N_{e} = 60 $ for numerical evaluations.
Note that the scale of the inflation is nearly fixed in metric formalism, while it decreases for large $\xi$ in the Palatini formalism.
On the other hand, the inflation ends at the field values of
\begin{align}  
\rho_{e} \simeq    \begin{dcases}
          \left(\frac{4}{3{\xi^{2}}}\right)^{1/4} {M_{P}} & (\text{metric}) \\  
         \left(\frac{8}{\xi}\right)^{1/4} M_{P} & (\text{Palatini})
        \end{dcases}.
\label{eq:endfield_values} 
\end{align}
in the large $\xi$ limit.

When identifying the $U(1)$ radial mode as the inflaton, its field value becomes time dependent and one should be careful which vev value of $ \rho $ to choose.
For instance, the isocurvature bound only probes the {field values} corresponding to the CMB pivot scale given in Eq.~\eqref{eq:field_values}. 
On the other hand, DM abundance receives contribution from the quantum fluctuation in various scale during its field excursion.
The largest contribution comes at the time of the end of inflation with $ \rho \simeq \rho_{e}$.
Therefore, for DM abundance from misalignment, we have to take $ \Omega_{a} = \Omega_{a} ( \langle \theta_{\text{mis},e}^{2} \rangle ) $ where $\langle \theta_{\text{mis}, e}^{2} \rangle$ has the vev of $ \rho $ field $ f_{a} $ replaced by $ \rho_{e} $, explicitly,
\begin{align}
    \langle \theta_{\text{mis},e}^{2} \rangle = \theta_{i}^2 + \left( \frac{H_{\rm inf}}{2\pi \rho_{e}} \right)^2 \,. \label{eq:theta_end}
\end{align}
The isocurvature fraction $\beta_{\rm iso} $ is determined at the CMB pivot scale where the corresponding field values are given in Eq.~\eqref{eq:field_values} as
\begin{align}
\beta_{\rm iso} \simeq \begin{dcases}
    \frac{9\xi}{16 \pi N_{e}^{3} \langle \theta_{\rm mis, *}^{2} \rangle }  { \left( \frac{\Omega_{a}}{\Omega_{\rm CDM} } \right)^{2} } \, , & (\text{metric}) \\
     \frac{1}{ 64 \pi N_{e}^{3} \xi \langle \theta_{\rm mis, *}^{2} \rangle } { \left( \frac{\Omega_{a}}{\Omega_{\rm CDM}} \right)^{2} } \, , & (\text{Palatini})
\end{dcases}    
\end{align}
where we have used Eqs.~\eqref{eq:isocurvature}, \eqref{eq:field_values}, and \eqref{eq:radial_Hinf}, and $\langle \theta_{\rm mis,*}^{2} \rangle$ is given by
\begin{align}
    \langle \theta_{\rm mis,*}^{2} \rangle = \theta_{i}^2 + \left( \frac{H_{\rm inf}}{2\pi \rho_{*}} \right)^2 \, .
\end{align}
We again note that $ \Omega_{a} $ is given with Eq.~\eqref{eq:theta_end} as discussed above.

\begin{figure*}
    \centering
    \includegraphics[width=0.45\textwidth]{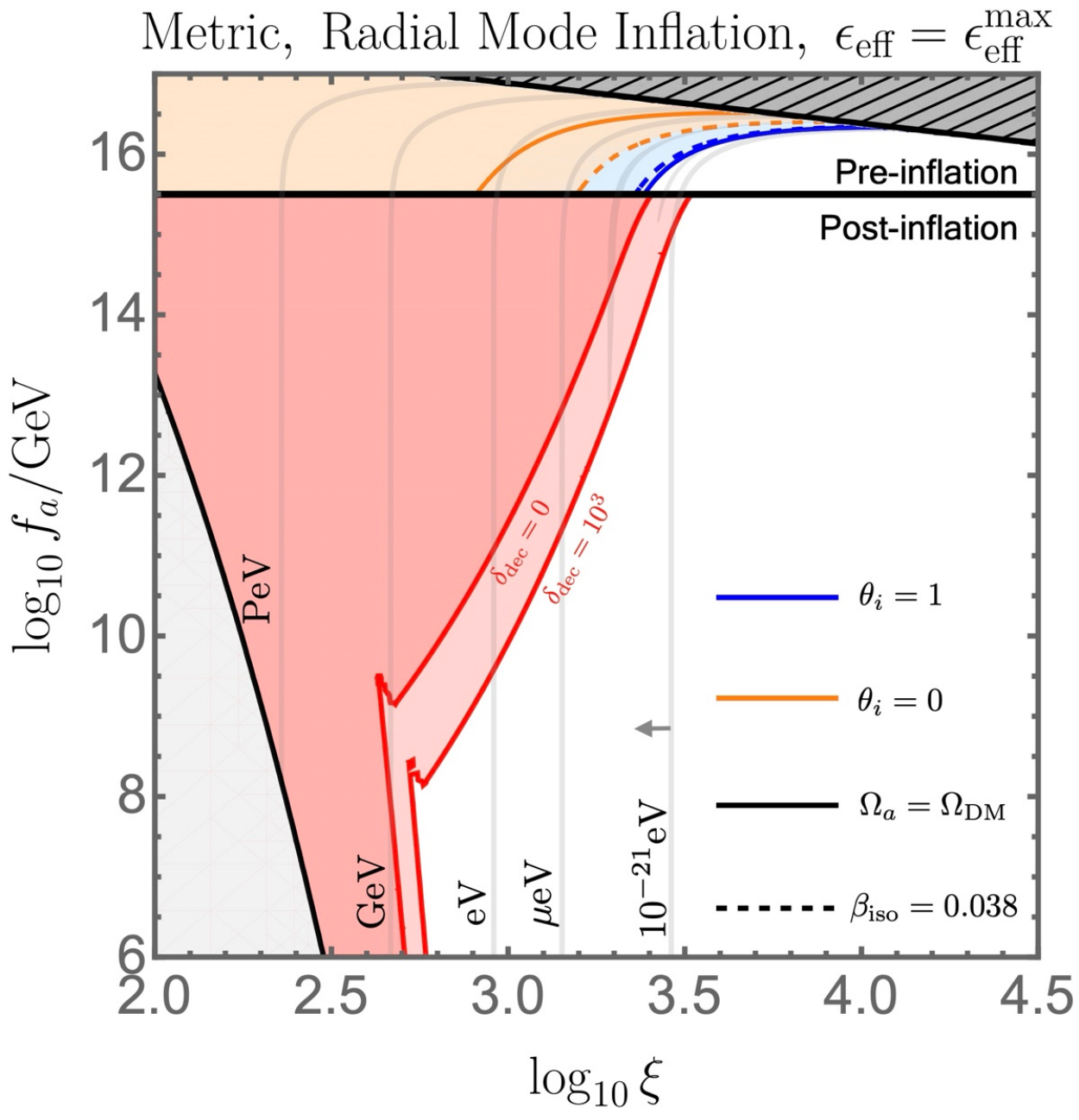}
    \includegraphics[width=0.45\textwidth]{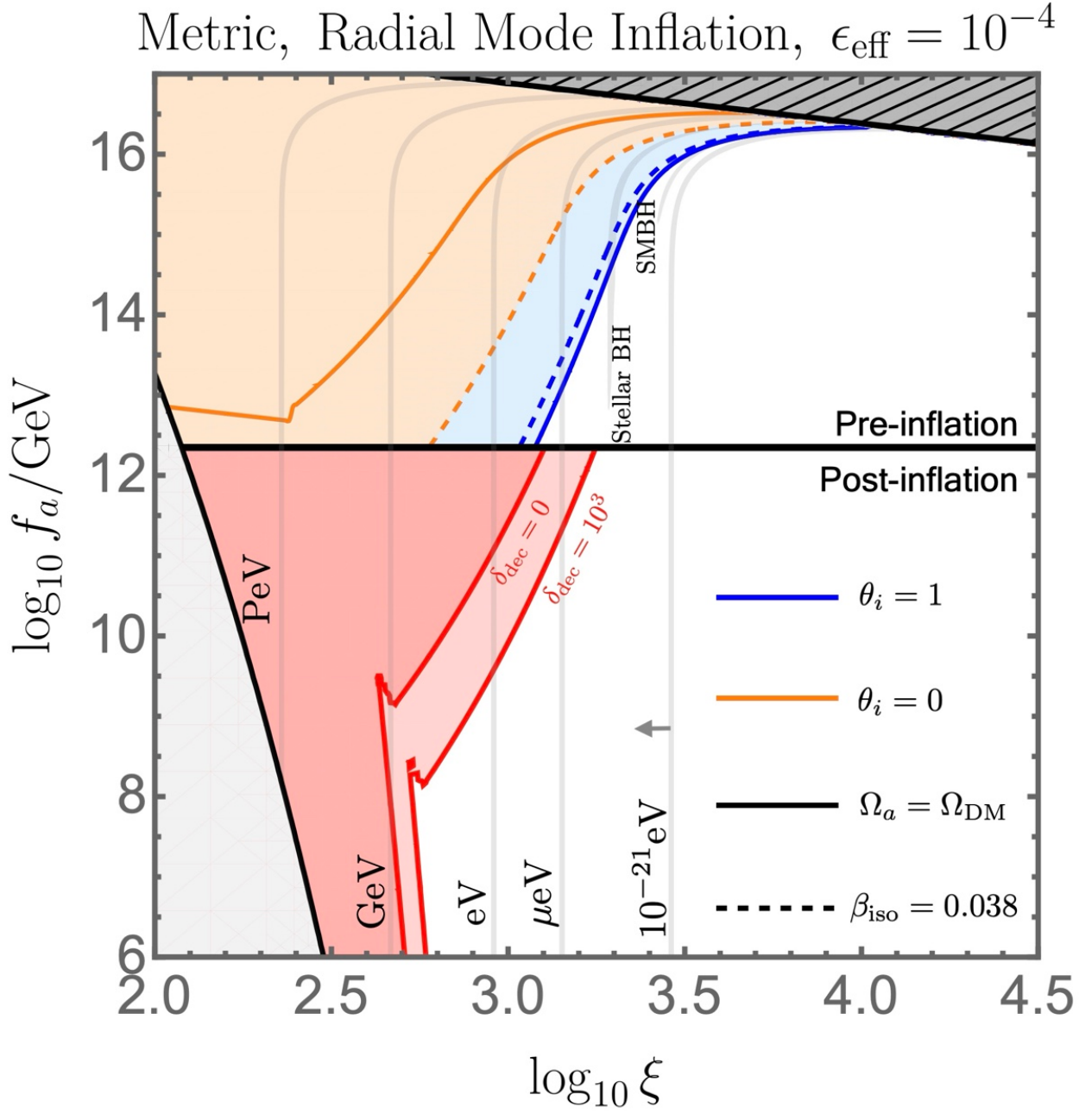}
    \includegraphics[width=0.45\textwidth]{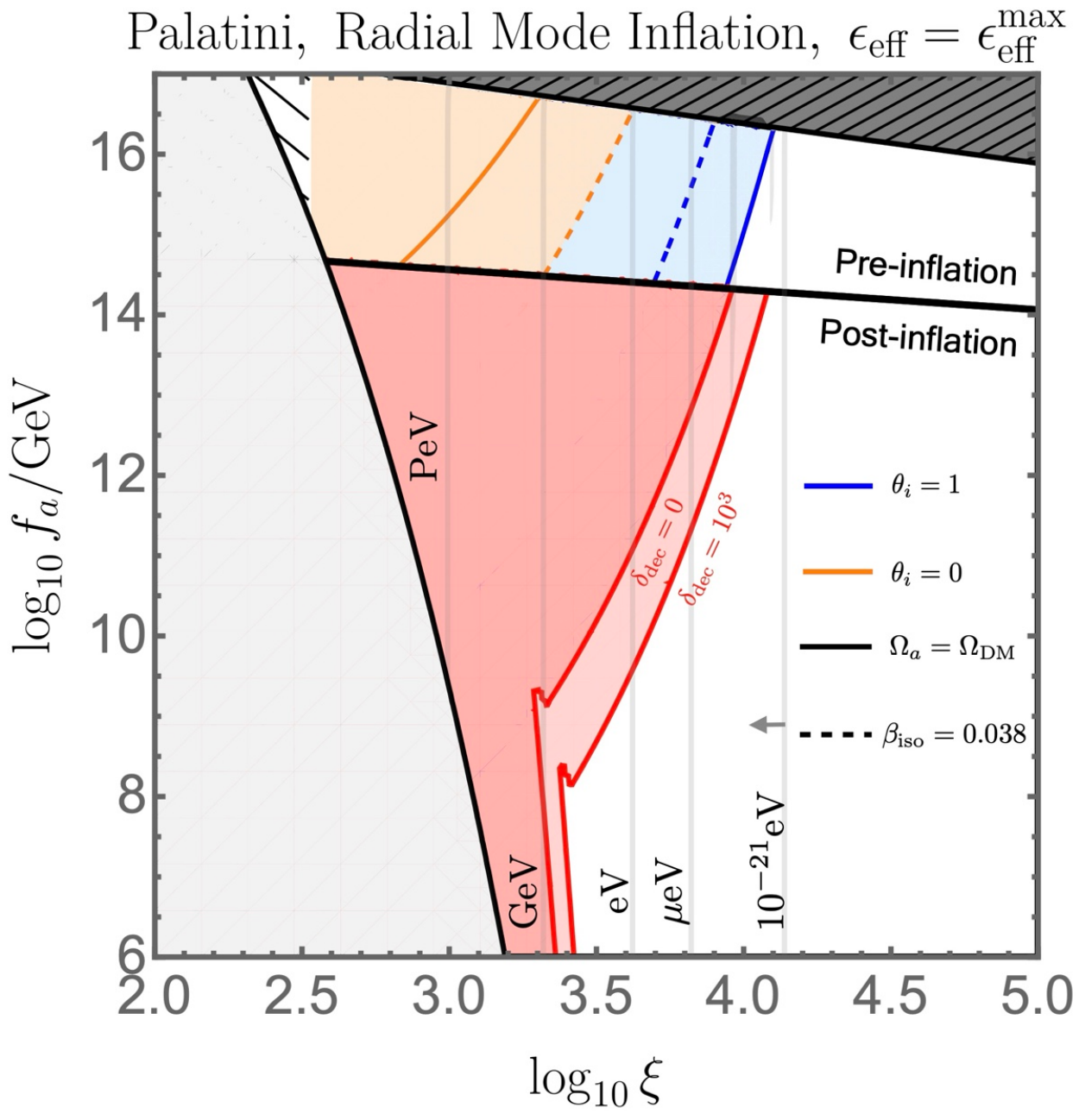}
    \includegraphics[width=0.45\textwidth]{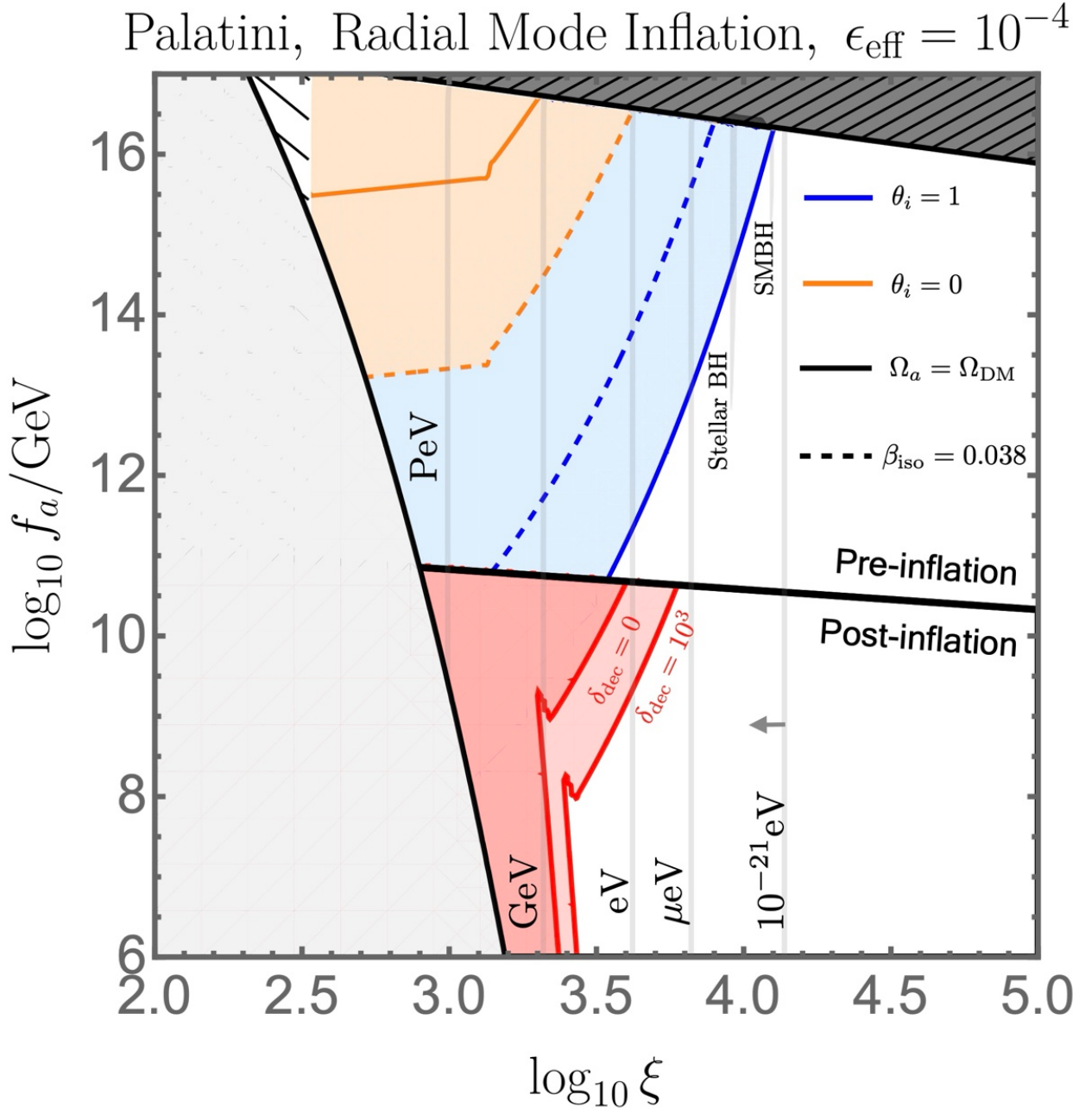}
    \caption{Constraints for the radial mode inflation 
    case for the metric formalism (top) and the Palatini formalism (bottom), with the same color scheme with Figure~\ref{fig:generic_metric_1}.
    The different dependencies of the constraints for $\theta_{i} = 0 $ originate from the different contributions in the quantum fluctuation part of $\langle \theta_{\rm mis,*}^2 \rangle $ and $\langle \theta_{\rm mis,e}^2 \rangle $. 
    The figures also incorporate $\lambda_{\Phi}$ values compatible with CMB normalization in Eq.~(\ref{eq:xiinf}), further widening the $m_{\rho} < m_{a} $ constraint region (light gray).
\label{fig:misalignment}
}\end{figure*}

Note that the $\xi$ appears in the denominator for Palatini case, which further suppresses the isocurvature perturbation in Palatini case. Taking those into account, the parameter regions are depicted in Figure~\ref{fig:misalignment}.
The constraints look schematically similar with general inflation cases, but exhibit several unique features.
First, as now $m_{a}^2 \sim (1/\rho_{e} L_{w}^3) \exp(-S_{w})$
at the end of the inflation,
this slightly alters the no-misalignment region. The region for $m_{\rho} < m_{a}$ also differs, as for radial mode inflation the $\lambda_{\Phi}$ value is determined through Eq.~(\ref{eq:xiinf}) following CMB normalization.
This makes a more notable change in the Palatini case as the required $\lambda_{\Phi}$ value (hence the corresponding $m_{\rho}$) is significantly smaller for lower $\xi$ compared to the metric case.
The different quantum fluctuation contributions for $\langle \theta_{\rm mis,*}^2 \rangle $ and $\langle \theta_{ \text{mis},e}^2 \rangle $ induce a $\xi$ dependence, which manifests in the $\theta_{i} = 0  $ case.
For large misalignment angles $ \theta_i = 1 $, we can see that the isocurvature constraints are always less stringent than the DM overabundance.
This allows the case for pre-inflationary ALP DM even for larger $H_{\rm inf}$ values, which is in contrast with the general inflation case.
For pre-inflation production the Palatini formalism allows a wider mass range, which is the opposite for the post-inflation production.

Let us discuss possible implications from the reheating. For a stable ALP, there could be lower limit on  the reheating temperature coming from the relativistic ALP remnant from the inflaton decay during the reheating stages \cite{Mazumdar:2016nzr,Kaneta:2024yyn}, which could be constrained by the non-observation of the dark radiation as
$ \Delta N_{\rm eff} \lesssim 0.3 $ \cite{Planck:2018jri}.
For instance, in the case when the ALP from the decaying radial mode inflaton is light enough to remain relativistic at the time of CMB or BBN, the reheating temperature $ T_{\rm reh}$ should be larger than $ 4 \times 10^{13} \, \text{GeV} $ if $ \xi \gtrsim 10^{4} $ and this lower bound becomes weaker for lower $\xi$ values for the case of $f_{a} < M_{P} / \xi $ \cite{Kaneta:2024yyn}.
In the case of massive ALP, there exists further complication of transition to the non-relativistic dark matter at late times while making the constraints from $ \Delta N_{\rm eff} $ further suppressed.
%

\section{Conclusion}
\label{section:conclusion}

In this work we considered the possibility of ALPs with their mass being solely gravitationally induced, and investigated the possibility of these consisting our dark matter. Non-perturbative gravitational effects on global symmetry contribute with an exponential dependence on the wormhole action. We obtained analytic results of the ALP wormhole solution for both metric and Palatini formalism, with both wormhole actions proportional to $\sqrt{\xi}$ for $\xi\gg 1$, and the metric formalism having a non-trivial $\xi$ dependence near the Giddings-Strominger limit. Identifying that the ALPs can take a large range of masses, we focused on the masses $m_{a} \gtrsim 10^{-21}~\mathrm{eV}$ for dark matter purposes.
We computed the ALP abundance for both pre-inflationary and post-inflationary setups, with the former coming from misalignment taking into account the period of reheating and the latter having additional contributions from cosmic string decays.
We further showed that constraints from isocurvature fluctuations differ when the radial mode $\rho$ plays the role of the inflaton,
and overall demonstrated that these ALPs can account for the observed dark matter density over a broad range of mass and symmetry breaking scales, driven purely by non-perturbative gravitational effects.

There are several further directions to consider within this scenario. In our analysis, we restricted the radial mode to be heavier than the ALP in order for it to stabilize at a vev.
Although this region is effectively constrained by ALP overproduction and isocurvature perturbations, it could be interesting to rigorously consider the case when this relation breaks down.

We also parameterized the reheating temperature without specifying a concrete example on the reheating process.
A more concrete setup could also induce interesting consequences. For example, if the complex $U(1)$ scalar field couples to right-handed neutrinos via $- \kappa_{ij} \Phi N_i N_j$, this allows the radial mode inflaton $\rho$ to decay into these right-handed neutrinos. This setup can then potentially explain the baryon asymmetry of our Universe through a non-thermal leptogenesis~\cite{Fukugita:1986hr, Lazarides:1990huy, Kumekawa:1994gx, Lazarides:1999dm, Giudice:1999fb, Asaka:1999yd, Asaka:1999jb} as well as neutrino mass through the seesaw mechanism~\cite{Minkowski:1977sc, Yanagida:1979as, Gell-Mann:1979vob, Mohapatra:1979ia}.
We leave a solid analysis on this scenario for a future work.

\section*{Acknowledgments}

The authors thank Guillermo Ballesteros, Heejoo Kim, and Chang Sub Shin for insightful discussions.
The authors further acknowledge the Workshop on Physics of Dark Cosmos held in Busan, Korea in 2022 where this project was initiated.
This work was supported by the National Research Foundation of Korea (NRF) grant funded by the Korea government (MSIT) (RS-2024-00340153) [DYC, SCP]. 
This work was supported by JSPS KAKENHI Grant Numbers 24H02244 (KH), 24K07041 (KH). 
The work of SML was supported by Samsung Science Technology Foundation under Project Number SSTF-BA2302-05 and the National Research Foundation of Korea (NRF) Grant RS-2023-00211732 and 2012K1A3A2A0105178151. 
The work of N.N. was supported in part by the Grant-in-Aid for Young Scientists (No.~21K13916).  
The work of SCP was further supported by the National Research Foundation of Korea (NRF) grant funded by the Korea government(MSIT)(RS-2023-00283129) and Yonsei internal grant for Mega-science (2023-22-048).


\bibliographystyle{JHEP}
\bibliography{ALP}

\providecommand{\href}[2]{#2}\begingroup\raggedright\begin{thebibliography}{10}

\bibitem{Peccei:1977hh}
R.D.~Peccei and H.R.~Quinn, \emph{{CP Conservation in the Presence of Instantons}}, \href{https://doi.org/10.1103/PhysRevLett.38.1440}{\emph{Phys. Rev. Lett.} {\bfseries 38} (1977) 1440}.

\bibitem{Peccei:1977ur}
R.D.~Peccei and H.R.~Quinn, \emph{{Constraints Imposed by CP Conservation in the Presence of Instantons}}, \href{https://doi.org/10.1103/PhysRevD.16.1791}{\emph{Phys. Rev. D} {\bfseries 16} (1977) 1791}.

\bibitem{Wilczek:1977pj}
F.~Wilczek, \emph{{Problem of Strong $P$ and $T$ Invariance in the Presence of Instantons}}, \href{https://doi.org/10.1103/PhysRevLett.40.279}{\emph{Phys. Rev. Lett.} {\bfseries 40} (1978) 279}.

\bibitem{Weinberg:1977ma}
S.~Weinberg, \emph{{A New Light Boson?}}, \href{https://doi.org/10.1103/PhysRevLett.40.223}{\emph{Phys. Rev. Lett.} {\bfseries 40} (1978) 223}.

\bibitem{Preskill:1982cy}
J.~Preskill, M.B.~Wise and F.~Wilczek, \emph{{Cosmology of the Invisible Axion}}, \href{https://doi.org/10.1016/0370-2693(83)90637-8}{\emph{Phys. Lett. B} {\bfseries 120} (1983) 127}.

\bibitem{Abbott:1982af}
L.F.~Abbott and P.~Sikivie, \emph{{A Cosmological Bound on the Invisible Axion}}, \href{https://doi.org/10.1016/0370-2693(83)90638-X}{\emph{Phys. Lett. B} {\bfseries 120} (1983) 133}.

\bibitem{Dine:1982ah}
M.~Dine and W.~Fischler, \emph{{The Not So Harmless Axion}}, \href{https://doi.org/10.1016/0370-2693(83)90639-1}{\emph{Phys. Lett. B} {\bfseries 120} (1983) 137}.

\bibitem{Kallosh:1995hi}
R.~Kallosh, A.D.~Linde, D.A.~Linde and L.~Susskind, \emph{{Gravity and global symmetries}}, \href{https://doi.org/10.1103/PhysRevD.52.912}{\emph{Phys. Rev. D} {\bfseries 52} (1995) 912} [\href{https://arxiv.org/abs/hep-th/9502069}{{\ttfamily hep-th/9502069}}].

\bibitem{Banks:2010zn}
T.~Banks and N.~Seiberg, \emph{{Symmetries and Strings in Field Theory and Gravity}}, \href{https://doi.org/10.1103/PhysRevD.83.084019}{\emph{Phys. Rev. D} {\bfseries 83} (2011) 084019} [\href{https://arxiv.org/abs/1011.5120}{{\ttfamily 1011.5120}}].

\bibitem{Witten:2017hdv}
E.~Witten, \emph{{Symmetry and Emergence}}, \href{https://doi.org/10.1038/nphys4348}{\emph{Nature Phys.} {\bfseries 14} (2018) 116} [\href{https://arxiv.org/abs/1710.01791}{{\ttfamily 1710.01791}}].

\bibitem{Harlow:2018jwu}
D.~Harlow and H.~Ooguri, \emph{{Constraints on Symmetries from Holography}}, \href{https://doi.org/10.1103/PhysRevLett.122.191601}{\emph{Phys. Rev. Lett.} {\bfseries 122} (2019) 191601} [\href{https://arxiv.org/abs/1810.05337}{{\ttfamily 1810.05337}}].

\bibitem{Harlow:2018tng}
D.~Harlow and H.~Ooguri, \emph{{Symmetries in quantum field theory and quantum gravity}}, \href{https://doi.org/10.1007/s00220-021-04040-y}{\emph{Commun. Math. Phys.} {\bfseries 383} (2021) 1669} [\href{https://arxiv.org/abs/1810.05338}{{\ttfamily 1810.05338}}].

\bibitem{Lee:1988ge}
K.-M.~Lee, \emph{{Wormholes and Goldstone Bosons}}, \href{https://doi.org/10.1103/PhysRevLett.61.263}{\emph{Phys. Rev. Lett.} {\bfseries 61} (1988) 263}.

\bibitem{Giddings:1989bq}
S.B.~Giddings and A.~Strominger, \emph{{STRING WORMHOLES}}, \href{https://doi.org/10.1016/0370-2693(89)91651-1}{\emph{Phys. Lett. B} {\bfseries 230} (1989) 46}.

\bibitem{Abbott:1989jw}
L.F.~Abbott and M.B.~Wise, \emph{{Wormholes and Global Symmetries}}, \href{https://doi.org/10.1016/0550-3213(89)90503-8}{\emph{Nucl. Phys. B} {\bfseries 325} (1989) 687}.

\bibitem{Coleman:1989zu}
S.R.~Coleman and K.-M.~Lee, \emph{{WORMHOLES MADE WITHOUT MASSLESS MATTER FIELDS}}, \href{https://doi.org/10.1016/0550-3213(90)90149-8}{\emph{Nucl. Phys. B} {\bfseries 329} (1990) 387}.

\bibitem{Hebecker:2016dsw}
A.~Hebecker, P.~Mangat, S.~Theisen and L.T.~Witkowski, \emph{{Can Gravitational Instantons Really Constrain Axion Inflation?}}, \href{https://doi.org/10.1007/JHEP02(2017)097}{\emph{JHEP} {\bfseries 02} (2017) 097} [\href{https://arxiv.org/abs/1607.06814}{{\ttfamily 1607.06814}}].

\bibitem{Alonso:2017avz}
R.~Alonso and A.~Urbano, \emph{{Wormholes and masses for Goldstone bosons}}, \href{https://doi.org/10.1007/JHEP02(2019)136}{\emph{JHEP} {\bfseries 02} (2019) 136} [\href{https://arxiv.org/abs/1706.07415}{{\ttfamily 1706.07415}}].

\bibitem{Hertog:2018kbz}
T.~Hertog, B.~Truijen and T.~Van~Riet, \emph{{Euclidean axion wormholes have multiple negative modes}}, \href{https://doi.org/10.1103/PhysRevLett.123.081302}{\emph{Phys. Rev. Lett.} {\bfseries 123} (2019) 081302} [\href{https://arxiv.org/abs/1811.12690}{{\ttfamily 1811.12690}}].

\bibitem{Hebecker:2018ofv}
A.~Hebecker, T.~Mikhail and P.~Soler, \emph{{Euclidean wormholes, baby universes, and their impact on particle physics and cosmology}}, \href{https://doi.org/10.3389/fspas.2018.00035}{\emph{Front. Astron. Space Sci.} {\bfseries 5} (2018) 35} [\href{https://arxiv.org/abs/1807.00824}{{\ttfamily 1807.00824}}].

\bibitem{Loges:2022nuw}
G.J.~Loges, G.~Shiu and N.~Sudhir, \emph{{Complex saddles and Euclidean wormholes in the Lorentzian path integral}}, \href{https://doi.org/10.1007/JHEP08(2022)064}{\emph{JHEP} {\bfseries 08} (2022) 064} [\href{https://arxiv.org/abs/2203.01956}{{\ttfamily 2203.01956}}].

\bibitem{Andriolo:2022rxc}
S.~Andriolo, G.~Shiu, P.~Soler and T.~Van~Riet, \emph{{Axion wormholes with massive dilaton}}, \href{https://doi.org/10.1088/1361-6382/ac8fdc}{\emph{Class. Quant. Grav.} {\bfseries 39} (2022) 215014} [\href{https://arxiv.org/abs/2205.01119}{{\ttfamily 2205.01119}}].

\bibitem{Loges:2023ypl}
G.J.~Loges, G.~Shiu and T.~Van~Riet, \emph{{A 10d construction of Euclidean axion wormholes in flat and AdS space}}, \href{https://doi.org/10.1007/JHEP06(2023)079}{\emph{JHEP} {\bfseries 06} (2023) 079} [\href{https://arxiv.org/abs/2302.03688}{{\ttfamily 2302.03688}}].

\bibitem{Jonas:2023ipa}
C.~Jonas, G.~Lavrelashvili and J.-L.~Lehners, \emph{{Zoo of axionic wormholes}}, \href{https://doi.org/10.1103/PhysRevD.108.066012}{\emph{Phys. Rev. D} {\bfseries 108} (2023) 066012} [\href{https://arxiv.org/abs/2306.11129}{{\ttfamily 2306.11129}}].

\bibitem{Kanazawa:2023xzy}
Y.~Kanazawa, \emph{{Axionic wormholes with R2 correction in metric and Palatini formulations}}, \href{https://doi.org/10.1103/PhysRevD.109.076009}{\emph{Phys. Rev. D} {\bfseries 109} (2024) 076009} [\href{https://arxiv.org/abs/2310.14641}{{\ttfamily 2310.14641}}].

\bibitem{Martucci:2024trp}
L.~Martucci, N.~Risso, A.~Valenti and L.~Vecchi, \emph{{Wormholes in the axiverse, and the species scale}}, \href{https://doi.org/10.1007/JHEP07(2024)240}{\emph{JHEP} {\bfseries 07} (2024) 240} [\href{https://arxiv.org/abs/2404.14489}{{\ttfamily 2404.14489}}].

\bibitem{Hertog:2024nys}
T.~Hertog, S.~Maenaut, B.~Missoni, R.~Tielemans and T.~Van~Riet, \emph{{Stability of Axion-Saxion wormholes}},  \href{https://arxiv.org/abs/2405.02072}{{\ttfamily 2405.02072}}.

\bibitem{Dine:1986bg}
M.~Dine and N.~Seiberg, \emph{{String Theory and the Strong {CP} Problem}}, \href{https://doi.org/10.1016/0550-3213(86)90043-X}{\emph{Nucl. Phys. B} {\bfseries 273} (1986) 109}.

\bibitem{Kamionkowski:1992mf}
M.~Kamionkowski and J.~March-Russell, \emph{{Planck scale physics and the Peccei-Quinn mechanism}}, \href{https://doi.org/10.1016/0370-2693(92)90492-M}{\emph{Phys. Lett. B} {\bfseries 282} (1992) 137} [\href{https://arxiv.org/abs/hep-th/9202003}{{\ttfamily hep-th/9202003}}].

\bibitem{Barr:1992qq}
S.M.~Barr and D.~Seckel, \emph{{Planck scale corrections to axion models}}, \href{https://doi.org/10.1103/PhysRevD.46.539}{\emph{Phys. Rev. D} {\bfseries 46} (1992) 539}.

\bibitem{Holman:1992us}
R.~Holman, S.D.H.~Hsu, T.W.~Kephart, E.W.~Kolb, R.~Watkins and L.M.~Widrow, \emph{{Solutions to the strong CP problem in a world with gravity}}, \href{https://doi.org/10.1016/0370-2693(92)90491-L}{\emph{Phys. Lett. B} {\bfseries 282} (1992) 132} [\href{https://arxiv.org/abs/hep-ph/9203206}{{\ttfamily hep-ph/9203206}}].

\bibitem{Ghigna:1992iv}
S.~Ghigna, M.~Lusignoli and M.~Roncadelli, \emph{{Instability of the invisible axion}}, \href{https://doi.org/10.1016/0370-2693(92)90019-Z}{\emph{Phys. Lett. B} {\bfseries 283} (1992) 278}.

\bibitem{Alvey:2020nyh}
J.~Alvey and M.~Escudero, \emph{{The axion quality problem: global symmetry breaking and wormholes}}, \href{https://doi.org/10.1007/JHEP01(2021)032}{\emph{JHEP} {\bfseries 01} (2021) 032} [\href{https://arxiv.org/abs/2009.03917}{{\ttfamily 2009.03917}}].

\bibitem{Sikivie:1982qv}
P.~Sikivie, \emph{{Of Axions, Domain Walls and the Early Universe}}, \href{https://doi.org/10.1103/PhysRevLett.48.1156}{\emph{Phys. Rev. Lett.} {\bfseries 48} (1982) 1156}.

\bibitem{Vilenkin:1984ib}
A.~Vilenkin, \emph{{Cosmic Strings and Domain Walls}}, \href{https://doi.org/10.1016/0370-1573(85)90033-X}{\emph{Phys. Rept.} {\bfseries 121} (1985) 263}.

\bibitem{Davis:1986xc}
R.L.~Davis, \emph{{Cosmic Axions from Cosmic Strings}}, \href{https://doi.org/10.1016/0370-2693(86)90300-X}{\emph{Phys. Lett. B} {\bfseries 180} (1986) 225}.

\bibitem{Vincent:1996rb}
G.R.~Vincent, M.~Hindmarsh and M.~Sakellariadou, \emph{{Scaling and small scale structure in cosmic string networks}}, \href{https://doi.org/10.1103/PhysRevD.56.637}{\emph{Phys. Rev. D} {\bfseries 56} (1997) 637} [\href{https://arxiv.org/abs/astro-ph/9612135}{{\ttfamily astro-ph/9612135}}].

\bibitem{Kawasaki:2014sqa}
M.~Kawasaki, K.~Saikawa and T.~Sekiguchi, \emph{{Axion dark matter from topological defects}}, \href{https://doi.org/10.1103/PhysRevD.91.065014}{\emph{Phys. Rev. D} {\bfseries 91} (2015) 065014} [\href{https://arxiv.org/abs/1412.0789}{{\ttfamily 1412.0789}}].

\bibitem{Vilenkin:1986ku}
A.~Vilenkin and T.~Vachaspati, \emph{{Radiation of Goldstone Bosons From Cosmic Strings}}, \href{https://doi.org/10.1103/PhysRevD.35.1138}{\emph{Phys. Rev. D} {\bfseries 35} (1987) 1138}.

\bibitem{Hamaguchi:2021mmt}
K.~Hamaguchi, Y.~Kanazawa and N.~Nagata, \emph{{Axion quality problem alleviated by nonminimal coupling to gravity}}, \href{https://doi.org/10.1103/PhysRevD.105.076008}{\emph{Phys. Rev. D} {\bfseries 105} (2022) 076008} [\href{https://arxiv.org/abs/2108.13245}{{\ttfamily 2108.13245}}].

\bibitem{Cheong:2022ikv}
D.Y.~Cheong, K.~Hamaguchi, Y.~Kanazawa, S.M.~Lee, N.~Nagata and S.C.~Park, \emph{{Axion quality problem and nonminimal gravitational coupling in the Palatini formulation}}, \href{https://doi.org/10.1103/PhysRevD.108.015007}{\emph{Phys. Rev. D} {\bfseries 108} (2023) 015007} [\href{https://arxiv.org/abs/2210.11330}{{\ttfamily 2210.11330}}].

\bibitem{Cheong:2023hrj}
D.Y.~Cheong, S.C.~Park and C.S.~Shin, \emph{{Effective theory approach for axion wormholes}}, \href{https://doi.org/10.1007/JHEP07(2024)039}{\emph{JHEP} {\bfseries 07} (2024) 039} [\href{https://arxiv.org/abs/2310.11260}{{\ttfamily 2310.11260}}].

\bibitem{Giddings:1987cg}
S.B.~Giddings and A.~Strominger, \emph{{Axion Induced Topology Change in Quantum Gravity and String Theory}}, \href{https://doi.org/10.1016/0550-3213(88)90446-4}{\emph{Nucl. Phys. B} {\bfseries 306} (1988) 890}.

\bibitem{Einstein:1925}
A.~Einstein, \emph{{Einheitliche feldtheorie von gravitation und elektrizit\"at}}, {\emph{Verlag der Koeniglich-Preussichen Akademie der Wissenschaften} {\bfseries {\bfseries 22}} (July, 1925) 414}.

\bibitem{Ferraris:1982}
M.~Ferraris, M.~Francaviglia and C.~Reina, \emph{{Variational formulation of general relativity from 1915 to 1925 ``palatini's method'' discovered by einstein in 1925}}, \href{https://doi.org/10.1007/BF00756060}{\emph{General Relativity and Gravitation} {\bfseries {\bfseries 14} no.~3} (Mar, 1982) 243}.

\bibitem{Ferraris:1992dx}
M.~Ferraris, M.~Francaviglia and I.~Volovich, \emph{{The Universality of vacuum Einstein equations with cosmological constant}}, \href{https://doi.org/10.1088/0264-9381/11/6/015}{\emph{Class. Quant. Grav.} {\bfseries 11} (1994) 1505} [\href{https://arxiv.org/abs/gr-qc/9303007}{{\ttfamily gr-qc/9303007}}].

\bibitem{Magnano:1993bd}
G.~Magnano and L.M.~Sokolowski, \emph{{On physical equivalence between nonlinear gravity theories and a general relativistic selfgravitating scalar field}}, \href{https://doi.org/10.1103/PhysRevD.50.5039}{\emph{Phys. Rev. D} {\bfseries 50} (1994) 5039} [\href{https://arxiv.org/abs/gr-qc/9312008}{{\ttfamily gr-qc/9312008}}].

\bibitem{York:1972sj}
J.W.~York, Jr., \emph{{Role of conformal three geometry in the dynamics of gravitation}}, \href{https://doi.org/10.1103/PhysRevLett.28.1082}{\emph{Phys. Rev. Lett.} {\bfseries 28} (1972) 1082}.

\bibitem{Gibbons:1976ue}
G.W.~Gibbons and S.W.~Hawking, \emph{{Action Integrals and Partition Functions in Quantum Gravity}}, \href{https://doi.org/10.1103/PhysRevD.15.2752}{\emph{Phys. Rev. D} {\bfseries 15} (1977) 2752}.

\bibitem{Ferreira:2020fam}
E.G.M.~Ferreira, \emph{{Ultra-light dark matter}}, \href{https://doi.org/10.1007/s00159-021-00135-6}{\emph{Astron. Astrophys. Rev.} {\bfseries 29} (2021) 7} [\href{https://arxiv.org/abs/2005.03254}{{\ttfamily 2005.03254}}].

\bibitem{Workman:2022ynf}
{\scshape Particle Data Group} collaboration, \emph{{Review of Particle Physics}}, \href{https://doi.org/10.1093/ptep/ptac097}{\emph{PTEP} {\bfseries 2022} (2022) 083C01}.

\bibitem{OHare:2024nmr}
C.A.J.~O'Hare, \emph{{Cosmology of axion dark matter}}, \href{https://doi.org/10.22323/1.454.0040}{\emph{PoS} {\bfseries COSMICWISPers} (2024) 040} [\href{https://arxiv.org/abs/2403.17697}{{\ttfamily 2403.17697}}].

\bibitem{AxionLimits}
C.~O'Hare, ``cajohare/axionlimits: Axionlimits.'' \url{https://cajohare.github.io/AxionLimits/}, July, 2020.
\newblock 10.5281/zenodo.3932430.

\bibitem{Stott:2018opm}
M.J.~Stott and D.J.E.~Marsh, \emph{{Black hole spin constraints on the mass spectrum and number of axionlike fields}}, \href{https://doi.org/10.1103/PhysRevD.98.083006}{\emph{Phys. Rev. D} {\bfseries 98} (2018) 083006} [\href{https://arxiv.org/abs/1805.02016}{{\ttfamily 1805.02016}}].

\bibitem{Baryakhtar:2020gao}
M.~Baryakhtar, M.~Galanis, R.~Lasenby and O.~Simon, \emph{{Black hole superradiance of self-interacting scalar fields}}, \href{https://doi.org/10.1103/PhysRevD.103.095019}{\emph{Phys. Rev. D} {\bfseries 103} (2021) 095019} [\href{https://arxiv.org/abs/2011.11646}{{\ttfamily 2011.11646}}].

\bibitem{Hoof:2024quk}
S.~Hoof, D.J.E.~Marsh, J.~Sisk-Reyn\'es, J.H.~Matthews and C.~Reynolds, \emph{{Getting More Out of Black Hole Superradiance: a Statistically Rigorous Approach to Ultralight Boson Constraints}},  \href{https://arxiv.org/abs/2406.10337}{{\ttfamily 2406.10337}}.

\bibitem{Kamionkowski:2022pkx}
M.~Kamionkowski and A.G.~Riess, \emph{{The Hubble Tension and Early Dark Energy}}, \href{https://doi.org/10.1146/annurev-nucl-111422-024107}{\emph{Ann. Rev. Nucl. Part. Sci.} {\bfseries 73} (2023) 153} [\href{https://arxiv.org/abs/2211.04492}{{\ttfamily 2211.04492}}].

\bibitem{Komatsu:2022nvu}
E.~Komatsu, \emph{{New physics from the polarized light of the cosmic microwave background}}, \href{https://doi.org/10.1038/s42254-022-00452-4}{\emph{Nature Rev. Phys.} {\bfseries 4} (2022) 452} [\href{https://arxiv.org/abs/2202.13919}{{\ttfamily 2202.13919}}].

\bibitem{DESI:2024mwx}
{\scshape DESI} collaboration, \emph{{DESI 2024 VI: Cosmological Constraints from the Measurements of Baryon Acoustic Oscillations}},  \href{https://arxiv.org/abs/2404.03002}{{\ttfamily 2404.03002}}.

\bibitem{ParticleDataGroup:2024cfk}
{\scshape Particle Data Group} collaboration, \emph{{Review of particle physics}}, \href{https://doi.org/10.1103/PhysRevD.110.030001}{\emph{Phys. Rev. D} {\bfseries 110} (2024) 030001}.

\bibitem{GrillidiCortona:2015jxo}
G.~Grilli~di Cortona, E.~Hardy, J.~Pardo~Vega and G.~Villadoro, \emph{{The QCD axion, precisely}}, \href{https://doi.org/10.1007/JHEP01(2016)034}{\emph{JHEP} {\bfseries 01} (2016) 034} [\href{https://arxiv.org/abs/1511.02867}{{\ttfamily 1511.02867}}].

\bibitem{Harari:1987ht}
D.~Harari and P.~Sikivie, \emph{{On the Evolution of Global Strings in the Early Universe}}, \href{https://doi.org/10.1016/0370-2693(87)90032-3}{\emph{Phys. Lett. B} {\bfseries 195} (1987) 361}.

\bibitem{Lyth:1991bb}
D.H.~Lyth, \emph{{Estimates of the cosmological axion density}}, \href{https://doi.org/10.1016/0370-2693(92)91590-6}{\emph{Phys. Lett. B} {\bfseries 275} (1992) 279}.

\bibitem{Klaer:2017qhr}
V.B.~Klaer and G.D.~Moore, \emph{{How to simulate global cosmic strings with large string tension}}, \href{https://doi.org/10.1088/1475-7516/2017/10/043}{\emph{JCAP} {\bfseries 10} (2017) 043} [\href{https://arxiv.org/abs/1707.05566}{{\ttfamily 1707.05566}}].

\bibitem{Gorghetto:2018myk}
M.~Gorghetto, E.~Hardy and G.~Villadoro, \emph{{Axions from Strings: the Attractive Solution}}, \href{https://doi.org/10.1007/JHEP07(2018)151}{\emph{JHEP} {\bfseries 07} (2018) 151} [\href{https://arxiv.org/abs/1806.04677}{{\ttfamily 1806.04677}}].

\bibitem{Kawasaki:2018bzv}
M.~Kawasaki, T.~Sekiguchi, M.~Yamaguchi and J.~Yokoyama, \emph{{Long-term dynamics of cosmological axion strings}}, \href{https://doi.org/10.1093/ptep/pty098}{\emph{PTEP} {\bfseries 2018} (2018) 091E01} [\href{https://arxiv.org/abs/1806.05566}{{\ttfamily 1806.05566}}].

\bibitem{Martins:2018dqg}
C.J.A.P.~Martins, \emph{{Scaling properties of cosmological axion strings}}, \href{https://doi.org/10.1016/j.physletb.2018.11.031}{\emph{Phys. Lett. B} {\bfseries 788} (2019) 147} [\href{https://arxiv.org/abs/1811.12678}{{\ttfamily 1811.12678}}].

\bibitem{Buschmann:2019icd}
M.~Buschmann, J.W.~Foster and B.R.~Safdi, \emph{{Early-Universe Simulations of the Cosmological Axion}}, \href{https://doi.org/10.1103/PhysRevLett.124.161103}{\emph{Phys. Rev. Lett.} {\bfseries 124} (2020) 161103} [\href{https://arxiv.org/abs/1906.00967}{{\ttfamily 1906.00967}}].

\bibitem{Hindmarsh:2019csc}
M.~Hindmarsh, J.~Lizarraga, A.~Lopez-Eiguren and J.~Urrestilla, \emph{{Scaling Density of Axion Strings}}, \href{https://doi.org/10.1103/PhysRevLett.124.021301}{\emph{Phys. Rev. Lett.} {\bfseries 124} (2020) 021301} [\href{https://arxiv.org/abs/1908.03522}{{\ttfamily 1908.03522}}].

\bibitem{Hook:2018dlk}
A.~Hook, \emph{{TASI Lectures on the Strong CP Problem and Axions}}, {\emph{PoS} {\bfseries TASI2018} (2019) 004} [\href{https://arxiv.org/abs/1812.02669}{{\ttfamily 1812.02669}}].

\bibitem{Benabou:2023ghl}
J.N.~Benabou, M.~Buschmann, S.~Kumar, Y.~Park and B.R.~Safdi, \emph{{Signatures of primordial energy injection from axion strings}}, \href{https://doi.org/10.1103/PhysRevD.109.055005}{\emph{Phys. Rev. D} {\bfseries 109} (2024) 055005} [\href{https://arxiv.org/abs/2308.01334}{{\ttfamily 2308.01334}}].

\bibitem{Drew:2023ptp}
A.~Drew, T.~Kinowski and E.P.S.~Shellard, \emph{{Axion string source modeling}}, \href{https://doi.org/10.1103/PhysRevD.110.043513}{\emph{Phys. Rev. D} {\bfseries 110} (2024) 043513} [\href{https://arxiv.org/abs/2312.07701}{{\ttfamily 2312.07701}}].

\bibitem{Benabou:2023npn}
J.N.~Benabou, Q.~Bonnefoy, M.~Buschmann, S.~Kumar and B.R.~Safdi, \emph{{Cosmological dynamics of string theory axion strings}}, \href{https://doi.org/10.1103/PhysRevD.110.035021}{\emph{Phys. Rev. D} {\bfseries 110} (2024) 035021} [\href{https://arxiv.org/abs/2312.08425}{{\ttfamily 2312.08425}}].

\bibitem{Saikawa:2024bta}
K.~Saikawa, J.~Redondo, A.~Vaquero and M.~Kaltschmidt, \emph{{Spectrum of global string networks and the axion dark matter mass}}, \href{https://doi.org/10.1088/1475-7516/2024/10/043}{\emph{JCAP} {\bfseries 10} (2024) 043} [\href{https://arxiv.org/abs/2401.17253}{{\ttfamily 2401.17253}}].

\bibitem{Kim:2024wku}
H.~Kim, J.~Park and M.~Son, \emph{{Axion dark matter from cosmic string network}}, \href{https://doi.org/10.1007/JHEP07(2024)150}{\emph{JHEP} {\bfseries 07} (2024) 150} [\href{https://arxiv.org/abs/2402.00741}{{\ttfamily 2402.00741}}].

\bibitem{Marsh:2015xka}
D.J.E.~Marsh, \emph{{Axion Cosmology}}, \href{https://doi.org/10.1016/j.physrep.2016.06.005}{\emph{Phys. Rept.} {\bfseries 643} (2016) 1} [\href{https://arxiv.org/abs/1510.07633}{{\ttfamily 1510.07633}}].

\bibitem{Kawasaki:2013ae}
M.~Kawasaki and K.~Nakayama, \emph{{Axions: Theory and Cosmological Role}}, \href{https://doi.org/10.1146/annurev-nucl-102212-170536}{\emph{Ann. Rev. Nucl. Part. Sci.} {\bfseries 63} (2013) 69} [\href{https://arxiv.org/abs/1301.1123}{{\ttfamily 1301.1123}}].

\bibitem{DiLuzio:2020wdo}
L.~Di~Luzio, M.~Giannotti, E.~Nardi and L.~Visinelli, \emph{{The landscape of QCD axion models}}, \href{https://doi.org/10.1016/j.physrep.2020.06.002}{\emph{Phys. Rept.} {\bfseries 870} (2020) 1} [\href{https://arxiv.org/abs/2003.01100}{{\ttfamily 2003.01100}}].

\bibitem{Hertzberg:2008wr}
M.P.~Hertzberg, M.~Tegmark and F.~Wilczek, \emph{{Axion Cosmology and the Energy Scale of Inflation}}, \href{https://doi.org/10.1103/PhysRevD.78.083507}{\emph{Phys. Rev. D} {\bfseries 78} (2008) 083507} [\href{https://arxiv.org/abs/0807.1726}{{\ttfamily 0807.1726}}].

\bibitem{Hamann:2009yf}
J.~Hamann, S.~Hannestad, G.G.~Raffelt and Y.Y.Y.~Wong, \emph{{Isocurvature forecast in the anthropic axion window}}, \href{https://doi.org/10.1088/1475-7516/2009/06/022}{\emph{JCAP} {\bfseries 06} (2009) 022} [\href{https://arxiv.org/abs/0904.0647}{{\ttfamily 0904.0647}}].

\bibitem{Beltran:2006sq}
M.~Beltran, J.~Garcia-Bellido and J.~Lesgourgues, \emph{{Isocurvature bounds on axions revisited}}, \href{https://doi.org/10.1103/PhysRevD.75.103507}{\emph{Phys. Rev. D} {\bfseries 75} (2007) 103507} [\href{https://arxiv.org/abs/hep-ph/0606107}{{\ttfamily hep-ph/0606107}}].

\bibitem{Fairbairn:2014zta}
M.~Fairbairn, R.~Hogan and D.J.E.~Marsh, \emph{{Unifying inflation and dark matter with the Peccei-Quinn field: observable axions and observable tensors}}, \href{https://doi.org/10.1103/PhysRevD.91.023509}{\emph{Phys. Rev. D} {\bfseries 91} (2015) 023509} [\href{https://arxiv.org/abs/1410.1752}{{\ttfamily 1410.1752}}].

\bibitem{Planck:2018jri}
{\scshape Planck} collaboration, \emph{{Planck 2018 results. X. Constraints on inflation}}, \href{https://doi.org/10.1051/0004-6361/201833887}{\emph{Astron. Astrophys.} {\bfseries 641} (2020) A10} [\href{https://arxiv.org/abs/1807.06211}{{\ttfamily 1807.06211}}].

\bibitem{Bezrukov:2007ep}
F.L.~Bezrukov and M.~Shaposhnikov, \emph{{The Standard Model Higgs boson as the inflaton}}, \href{https://doi.org/10.1016/j.physletb.2007.11.072}{\emph{Phys. Lett. B} {\bfseries 659} (2008) 703} [\href{https://arxiv.org/abs/0710.3755}{{\ttfamily 0710.3755}}].

\bibitem{Bauer:2008zj}
F.~Bauer and D.A.~Demir, \emph{{Inflation with Non-Minimal Coupling: Metric versus Palatini Formulations}}, \href{https://doi.org/10.1016/j.physletb.2008.06.014}{\emph{Phys. Lett. B} {\bfseries 665} (2008) 222} [\href{https://arxiv.org/abs/0803.2664}{{\ttfamily 0803.2664}}].

\bibitem{Cheong:2021kyc}
D.Y.~Cheong, S.M.~Lee and S.C.~Park, \emph{{Reheating in models with non-minimal coupling in metric and~Palatini formalisms}}, \href{https://doi.org/10.1088/1475-7516/2022/02/029}{\emph{JCAP} {\bfseries 02} (2022) 029} [\href{https://arxiv.org/abs/2111.00825}{{\ttfamily 2111.00825}}].

\bibitem{Mazumdar:2016nzr}
A.~Mazumdar, S.~Qutub and K.~Saikawa, \emph{{Nonthermal axion dark radiation and constraints}}, \href{https://doi.org/10.1103/PhysRevD.94.065030}{\emph{Phys. Rev. D} {\bfseries 94} (2016) 065030} [\href{https://arxiv.org/abs/1607.06958}{{\ttfamily 1607.06958}}].

\bibitem{Kaneta:2024yyn}
K.~Kaneta, S.M.~Lee, K.-y.~Oda and T.~Takahashi, \emph{{Pseudo-Nambu-Goldstone Boson Production from Inflaton Coupling during Reheating}},  \href{https://arxiv.org/abs/2406.09045}{{\ttfamily 2406.09045}}.

\bibitem{Fukugita:1986hr}
M.~Fukugita and T.~Yanagida, \emph{{Baryogenesis Without Grand Unification}}, \href{https://doi.org/10.1016/0370-2693(86)91126-3}{\emph{Phys. Lett. B} {\bfseries 174} (1986) 45}.

\bibitem{Lazarides:1990huy}
G.~Lazarides and Q.~Shafi, \emph{{Origin of matter in the inflationary cosmology}}, \href{https://doi.org/10.1016/0370-2693(91)91090-I}{\emph{Phys. Lett. B} {\bfseries 258} (1991) 305}.

\bibitem{Kumekawa:1994gx}
K.~Kumekawa, T.~Moroi and T.~Yanagida, \emph{{Flat potential for inflaton with a discrete R invariance in supergravity}}, \href{https://doi.org/10.1143/PTP.92.437}{\emph{Prog. Theor. Phys.} {\bfseries 92} (1994) 437} [\href{https://arxiv.org/abs/hep-ph/9405337}{{\ttfamily hep-ph/9405337}}].

\bibitem{Lazarides:1999dm}
G.~Lazarides, \emph{{Leptogenesis in supersymmetric hybrid inflation}}, {\emph{Springer Tracts Mod. Phys.} {\bfseries 163} (2000) 227} [\href{https://arxiv.org/abs/hep-ph/9904428}{{\ttfamily hep-ph/9904428}}].

\bibitem{Giudice:1999fb}
G.F.~Giudice, M.~Peloso, A.~Riotto and I.~Tkachev, \emph{{Production of massive fermions at preheating and leptogenesis}}, \href{https://doi.org/10.1088/1126-6708/1999/08/014}{\emph{JHEP} {\bfseries 08} (1999) 014} [\href{https://arxiv.org/abs/hep-ph/9905242}{{\ttfamily hep-ph/9905242}}].

\bibitem{Asaka:1999yd}
T.~Asaka, K.~Hamaguchi, M.~Kawasaki and T.~Yanagida, \emph{{Leptogenesis in inflaton decay}}, \href{https://doi.org/10.1016/S0370-2693(99)01020-5}{\emph{Phys. Lett. B} {\bfseries 464} (1999) 12} [\href{https://arxiv.org/abs/hep-ph/9906366}{{\ttfamily hep-ph/9906366}}].

\bibitem{Asaka:1999jb}
T.~Asaka, K.~Hamaguchi, M.~Kawasaki and T.~Yanagida, \emph{{Leptogenesis in inflationary universe}}, \href{https://doi.org/10.1103/PhysRevD.61.083512}{\emph{Phys. Rev. D} {\bfseries 61} (2000) 083512} [\href{https://arxiv.org/abs/hep-ph/9907559}{{\ttfamily hep-ph/9907559}}].

\bibitem{Minkowski:1977sc}
P.~Minkowski, \emph{{$\mu \to e\gamma$ at a Rate of One Out of $10^{9}$ Muon Decays?}}, \href{https://doi.org/10.1016/0370-2693(77)90435-X}{\emph{Phys. Lett. B} {\bfseries 67} (1977) 421}.

\bibitem{Yanagida:1979as}
T.~Yanagida, \emph{{Horizontal gauge symmetry and masses of neutrinos}}, {\emph{Conf. Proc. C} {\bfseries 7902131} (1979) 95}.

\bibitem{Gell-Mann:1979vob}
M.~Gell-Mann, P.~Ramond and R.~Slansky, \emph{{Complex Spinors and Unified Theories}}, {\emph{Conf. Proc. C} {\bfseries 790927} (1979) 315} [\href{https://arxiv.org/abs/1306.4669}{{\ttfamily 1306.4669}}].

\bibitem{Mohapatra:1979ia}
R.N.~Mohapatra and G.~Senjanovic, \emph{{Neutrino Mass and Spontaneous Parity Nonconservation}}, \href{https://doi.org/10.1103/PhysRevLett.44.912}{\emph{Phys. Rev. Lett.} {\bfseries 44} (1980) 912}.

\end{thebibliography}\endgroup

\end{document}